\documentclass[
twocolumn,aps,prx,showpacs
]{revtex4}

\usepackage{graphicx,color,hyperref}

\usepackage{amsmath}
\usepackage{dsfont}
\usepackage{amsfonts}
\usepackage{amssymb}

\newcommand{\tr}{\mathrm{tr}}
\newcommand{\Str}{\mathrm{Str}}
\newcommand{\STr}{\mathrm{STr}}

\begin{document}

\title{
A minimal model of many body localization}

\author{
F.~Monteiro, T.~Micklitz
}
\affiliation{
Centro Brasileiro de Pesquisas F\'isicas, Rua Xavier Sigaud 150, 22290-180, Rio de Janeiro, Brazil }

\author{Masaki Tezuka}
\affiliation{
Department of Physics, Kyoto University, Kyoto 606-8502, Japan
}

\author{Alexander Altland}
\affiliation{Institut f\"ur Theoretische Physik, Universit\"at zu K\"oln, Z\"ulpicher Str. 77, 50937 Cologne, Germany
}

\date{\today}

\pacs{05.45.Mt, 72.15.Rn, 71.30.+h}

\begin{abstract} 
We present a fully analytical description of a many body localization (MBL)
transition in a microscopically defined model. Its Hamiltonian is the sum of one-
and two-body operators, where both contributions obey a maximum-entropy principle and
have no symmetries except hermiticity (not even particle number conservation). These
two criteria paraphrase that our system is a variant of the Sachdev-Ye-Kitaev (SYK)
model. We will demonstrate how this simple `zero-dimensional' system displays
numerous features seen in more complex realizations of MBL. Specifically, it shows a
transition between an ergodic and a localized phase, and non-trivial wave function
statistics indicating the presence of `non-ergodic extended states'. We check our
analytical description of these phenomena by parameter free comparison to high
performance numerics for systems of up to $N=15$ fermions. In this way, our study
becomes a testbed for concepts of high-dimensional quantum localization, previously
applied to synthetic systems such as Cayley trees or random regular graphs. We
believe that this is the first many body system for which an effective theory is
derived and solved from first principles. The hope is that the novel analytical
concepts developed in this study may become a stepping stone for the description of
MBL in more complex systems.
\end{abstract}

\maketitle


\section{Introduction}

Quantum wave functions subject to strong static randomness may show non-ergodic
localized behavior. To date, we distinguish between two broad universality classes of
quantum localization: \emph{Anderson localization}~\cite{Anderson} in low
dimensional single particle systems, and \emph{many body localization} (MBL) in
random many particle systems~\cite{BaskoAleinerAltshuler,Mirlin}. 
In principle, there is no fundamental distinction
between these two. They both reflect the lack of ergodicity of wave functions 
on random lattices due to massive quantum interference. However, the all important difference is that in
the former case the lattice structure is defined by a low dimensional solid 
and in the latter by the high dimensional Fock space lattice formed by the occupation number states
of a many particle system. 

Many body localization is traditionally discussed in the context of spatially
extended many body systems, such as interacting quasi one-dimensional electron
systems~\cite{BaskoAleinerAltshuler,Mirlin}, or 
random spin chains~\cite{spin1,Huse2010,Moore2012,spin5,spin3,spin2,ImbriePRL2016,ImbrieJStatPhys2016,TorresSantos17,Laflorencie19}.
However, that spatial extension is an added layer of complexity to a problem that
manifests itself already in spatially confined geometries --- a competition between
`hopping' and `randomness' on the complex lattice structure defined by an interacting
particle problem. In fact, there appears to be a paradigm shift in the field away
from studying the quantum critical phenomena of the localization transition in
extended systems towards manifestations of MBL in systems of mesoscopic extension, such as interacting quantum
dots~\cite{AltshulerGefenKamenevLevitov97,Silvestrov97,Silvestrov98,GornyMirlinPolyakov16,GornyMirlinPolyakovBurin17},
small sized optical lattices~\cite{RubioAbadal19,Choi16,Schreiber15}, 
or small sized superconducting qubit arrays~\cite{kai18, roushan17}.
This development is driven in part by pragmatism. The explosion of Fock space dimensions with increasing system size makes numerical access infamously hard and classical computers
may never be powerful enough to probe the scaling regime of the MBL transition with
sufficient reliability. Another motivation lies in the fascinating and only partly
understood physics of localization in many body systems of
intermediate size. 

At this point, even basic aspects of MBL remain enigmatic, including in small sized
systems. Among these, one of the most controversial topics concerns the
presence or absence of a phase of non-ergodic but extended (NEE) states intermediate
between the regime of ergodic wave functions at weak and localized wave functions at
strong disorder. If existent, such a phase must be born out of the main principles
distinguishing MBL from low dimensional AL: the high coordination number of `Fock
space lattices', the strong correlation of their disorder potentials, and the
sparsity of the hopping matrix elements in Fock space (see the next section for a
more detailed discussion). Reflecting the complexity of the problem, the physics of
NEE states is often discussed for
synthetic~\cite{NonE_Ex,Biroli18,Mirlin1,TikhonovMirlin19_2,faoro} or
phenomenological models~\cite{RPmodel1}, sidestepping one or several of the above
complications. (However, even for these, the existence of NEE phases is discussed
controversially.)

Clearly, a numerically and analytically solvable minimal model  defined by a
\emph{microscopic}  Hamiltonian  would provide an important contribution to our
understanding of MBL. It  would provide a testbed for the validity of
analytical approaches by comparison to numerical diagonalization and might turn into
a building block in the study of more complex systems. In this paper, we report on the
definition and solution of such a system.
Here, the term `solution' refers to the following: (a) the construction of an \emph{effective} theory of the microscopically defined system by parametrically controlled approximation, (b) the computation of observables (many body wave function, and spectral statistics) from that theory, and (c) parameter free comparison to numerics. 
In this hierarchy, the perhaps most important element is (a). The effective theory we derive assumes the form of a matrix-path integral in Fock space, see Eq.~\eqref{QActionHopping} for an impression. From this representation, observables can be extracted  by powerful methods developed in the localization theory of high dimensional lattices.  (For a pioneering previous comparison between  analytical and numerical results for a concrete model system we refer to  Ref.~\cite{LeyronasSilvestrovBeenakker2000}. However, that work was based on scaling theory for a specific class of observables. Lacking element (a), it did not have the scope of the present analysis.)

The model we consider is implicitly defined by the
following criteria: its Hamiltonian $\hat H=\hat H_2 + \hat H_4$ contains the sum of
a one-body and a two-body part. Both are maximally entropic and have no symmetries
besides hermiticity (not even particle number conservation). In the non-interacting
case, $\hat H_4=0$, the product eigenstates of $\hat H_2$ define a basis in which the
system is trivially localized. The Hamiltonian $\hat H_4$ acts as a `hopping
operator' and at a critical strength will induce a many body localization transition.
In a manner detailed in the next section, the criteria listed above state that $\hat
H$ is the Majorana SYK Hamiltonian.

The maximum entropy criterion makes the SYK model much simpler than MBL systems with
spatial extension. At the same time, it displays a wealth of phenomena characteristic for MBL. Foremost among these is a change from delocalized to localized behavior. For finite $N$, this is a crossover. However, the exponential dependence of the Fock space lattice extension on $N$ implies that it rapidly acquires signatures of a transition as $N$ increases.  
Second, the model supports a regime (not a phase) of NEE states
prior to the onset of localization. We will discuss how the diminishing support of
these states upon approaching the transition reflects the structure of the system's
Fock space, and how this differs from phenomenological models. However, the most
important point of all is that the spectral and wave function statistics of the model
can be computed analytically and that these results can be numerically tested in a
parameter free comparison. The analytical approach is based on matrix integral
techniques imported from the theory of high dimensional random lattices.  We apply
these techniques subject to a number of assumptions which should generalize to other
many body systems of small spatial extension and/or a high degree of connectivity. 
We, therefore, hope that the approach discussed in this paper
may become a stepping stone for the solution of more complex manifestations of MBL.

\noindent \emph{Plan of the paper:} In the next section, we introduce our model
system, qualitatively discuss its physics, and summarize our main results. The
remaining parts of the paper discuss the derivation of these findings, where we try
to keep the technical level at a bare minimum. In Section~\ref{M_model} we map the
computation of disorder averaged correlation functions onto that of an equivalent
matrix integral. In Section~\ref{Effective_T}, a stationary phase approach is applied
to reduce the matrix integral to an effective theory describing physics at large time
scales. In sections~\ref{sec:wave_function_statistics} and~\ref{Ergodic_loc} we apply
this representation to the discussion of wave function statistics and the
localization transition, respectively. We conclude in Section~\ref{discussion} with a
discussion comparing our results to those obtained for other models, and on possible
generalizations to other MBL systems. Technical parts of our analysis are relegated
to a number of appendices.

\section{Model and summary of results} 
\label{sec:model_and_summary_of_results}

In this section, we first introduce the SYK model and then discuss its physics of quantum localization in qualitative terms. Much of this outline is formulated in general terms which should carry over to similar models. In the remaining parts of the section we get more concrete and summarize our results in comparison to numerics. 

\subsubsection{SYK Model} 
\label{ssub:syk_model}

The SYK Hamiltonian~\cite{SYK1,SYK2} 
\begin{align}
\label{syk}
\hat{H}_4
&=
{1\over 4!}\sum_{i,j,k,l=1}^{2N}
J_{ijkl} \hat{\chi}_i \hat{\chi}_j \hat{\chi}_k \hat{\chi}_l ,
\end{align}
describes a system of $2N$ Majorana fermions,
$\{\hat{\chi}_i,\hat{\chi}_j\}=2\delta_{ij}$, subject to an all-to-all interaction,
with matrix elements $\{J_{ijkl} \}$ drawn from a Gaussian distribution of variance
$\langle |J_{ijkl}|^2 \rangle= 6J^2/(2N)^3$. Defined in this way, it defines an ideal
of a massively interacting quantum system lacking any degree of internal structure.
Due to the `least information' principle realized through the stochastic interaction,
all single particle orbitals, $i$, stand on equal footing, and the absence of a
continuous $\mathrm{U}(1)$-symmetry prevents the fragmentation of the Fock space
into sectors of conserved particle number. Reflecting these features, the physics of
the SYK Hamiltonian at large time scales becomes equivalent to that of random matrix
theory (RMT), with wave functions homogeneously distributed over the full Hilbert
space.

A tendency to Fock space localization is included by adding to $\hat H_4$ a free particle contribution~\cite{SYK_GG,Shepelyansky17},
\begin{align}
\label{syk2chi}
\hat H_2 = \frac{1}{2}\sum_{i,j=1}^{2N}J_{ij}  \hat{\chi}_i  \hat{\chi}_j,
\end{align}
with a likewise random antisymmetric matrix $J_{ij}=-J_{ji}$, with matrix elements
$\{J_{ij} \}$ drawn from a Gaussian of variance $\langle |J_{ij}|^2 \rangle=
\delta^2/2N$. Without loss of generality, we may assume $\{J_{ij}\}$ to be
diagonalized into a form $\hat H_2=i\sum_{i}^{N}  v_i   \hat{\chi}_{2i-1}
\hat{\chi}_{2i}$, where $\pm  v_{i}$ are the eigenvalues of the hermitian matrix $i
\{J_{ij}\}$. For the above distribution of the matrix elements $J_{ij}$ 
these eigenvalues are random numbers with variance $\sim \delta$.

\begin{figure}
\centering
\includegraphics[width=8.5cm]{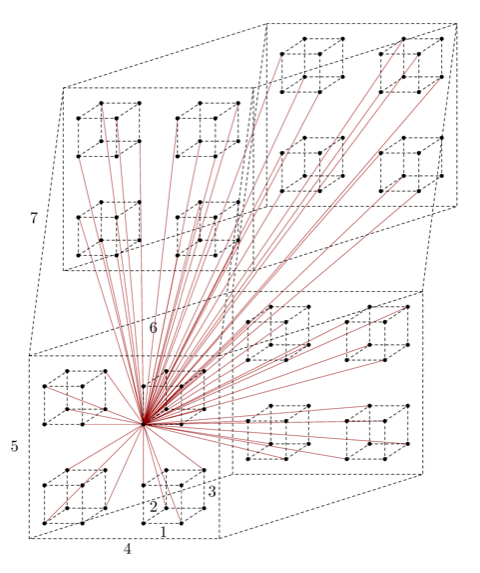}
\caption{\label{FockSpace} 
Hypercubical Fock space of a $2N=14$ Majorana system. The numbers indicate the bit depth of states in the computational fermion basis, and the lines are a qualitative representation of the connectivity of the reference state $|0, 0, 0, 1, 1, 0, 0\rangle$. For large values of $N$, the pattern of connections becomes sparse. However, there remain exponentially many $\propto D$ connections, statistically correlated due to the small number of $\sim N^4$ of independent random amplitudes.
 }
\end{figure}

We next translate from the Majorana many body Hamiltonian formulation   to one in
terms of a fermion Fock space (lattice). To this end, define $N$ complex fermion
annihilation operators $\hat{c}_i =
\tfrac{1}{2}(\hat{\chi}_{2i-1} + i\hat{\chi}_{2i})$ satisfying
$\{\hat{c}_{i}, \hat{c}_{j}^{\dagger}\} =
  \delta_{ij}$. With the  
number operators $\hat{n}_{i} = \hat{c}_{i}^{\dagger}\hat{c}_{i}$, we then have 
\begin{align}
\label{syk2n}
    \hat{H}_2 = \sum_{i=1}^N v_i (2\hat{n}_i - 1),\qquad\text{var}(v_i)=\delta^2.
\end{align}
Representing this Hamiltonian in the basis of  $2^N$ occupation number states, $
|n\rangle = |n_1,n_2,...,n_N\rangle$, $n_i=0,1$, it assumes the form of a random
potential $v_n=\sum_i v_i (2n_i-1)$ on the hypercube defined by all sites
$n=(\dots,0,0,1,0,0,1,0,\ldots)$~\cite{fn2}. In the same basis, the interaction, $\hat H_4$,
assumes the role of a fermion number conserving `hopping operator' $\hat H_4$
connecting sites of bit separation 2 and 4~\cite{fn1}. This hopping introduces a
complex connectivity pattern on the two decoupled sublattices of definite (even, say)
parity, containing
 \begin{align}
   \label{eq:DDef}
   D=2^{N-1}
 \end{align}
sites each. Fig.~\ref{FockSpace} illustrates this structure for a Fock space of $14$ Majorana fermions. The lines indicate the states connected to the arbitrarily chosen site
$|0, 0, 0, 1, 1, 0, 0\rangle$. Notice the  high coordination number and the absence of lattice
periodicity, symptomatic for this and for other Fock space lattices. 
  The competition between the localizing random potential $\hat H_2$
  and the delocalizing hopping $\hat H_4$
  defines the MBL problem,
  regardless of their detailed realization.

\subsection{Qualitative discussion} 
\label{ssub:qualitative_discussion}

In this section, we discuss the physics of the above random system in qualitative terms. Specific topics include the existence of a localization-delocalization transition, its signatures in spectral and wave function statistics, and a regime of non-ergodically extended states. Most parts of this discussion do not make  specific reference to the  SYK model and should equally apply to other systems. 

The single most important system quantity relevant to the understanding of the above observables at a specific energy, 
say, $E$, is the local density of states in Fock space,  
\begin{align}
    \label{LocalDoSDef}
    \nu_n \equiv -\frac{1}{\pi}\mathrm{Im}\left\langle \big\langle n \big| \frac{1}{E^+-\hat H_2 -\hat H_4}\big| n \big\rangle \right\rangle_J,
\end{align}
where $E^+\equiv E+i\epsilon$, and 
$\langle \cdots \rangle_J$ indicates that we consider
$\nu_n$ averaged over realizations of $\hat H_4$, but at a single realization of $\hat
H_2$. (The discussion above shows that the large coordination number of the lattice
makes $\nu_n$ a largely self-averaging quantity. Averaging over $\hat H_4$ is largely
a matter of technical convenience.) From the perspective of site $n$, the large
number of nearest neighbors represents an environment and, on this basis, one expects a
Lorentzian profile
\begin{align}
    \label{LocalDosLorentzian}
    \nu_n = \frac{1}{\pi}\frac{\kappa_n}{v_n^2+\kappa_n^2},
\end{align}
where we have set $E=0$ for definiteness, and the broadening $\kappa_n=\kappa_n(\Delta_4,\delta,\alpha)$ 
must be self-consistently determined (cf. Eq.~\eqref{SaddlePointEquation} below) in dependence on the following parameters:
\begin{itemize}
    \item the many body band width, $\Delta_4$, of the interaction operator ($\Delta_4=\sqrt{J^{2}N/2}$~\cite{fn3} for the SYK Hamiltonian, $\hat H_4$),
    \item the disorder strength, $\delta$, or, equivalently, the distribution width, $\Delta_2$ of the on-site random potential Eq.~\eqref{syk2n}. For large $N$, the central limit theorem implies $\Delta_2=\delta N^{1/2}$~\cite{fn3a}.
    \item The number $\sim N^\alpha$ of nearest neighbors, $m$,  connected to Fock space sites, $n$, by the interaction $\hat H_4$. ($\alpha=4$ for the SYK Hamiltonian.)
\end{itemize}
On this basis, we must distinguish between four regimes of qualitatively different level hybridization, $\kappa$:

\begin{figure}
\centering
\includegraphics[width=8.5cm]{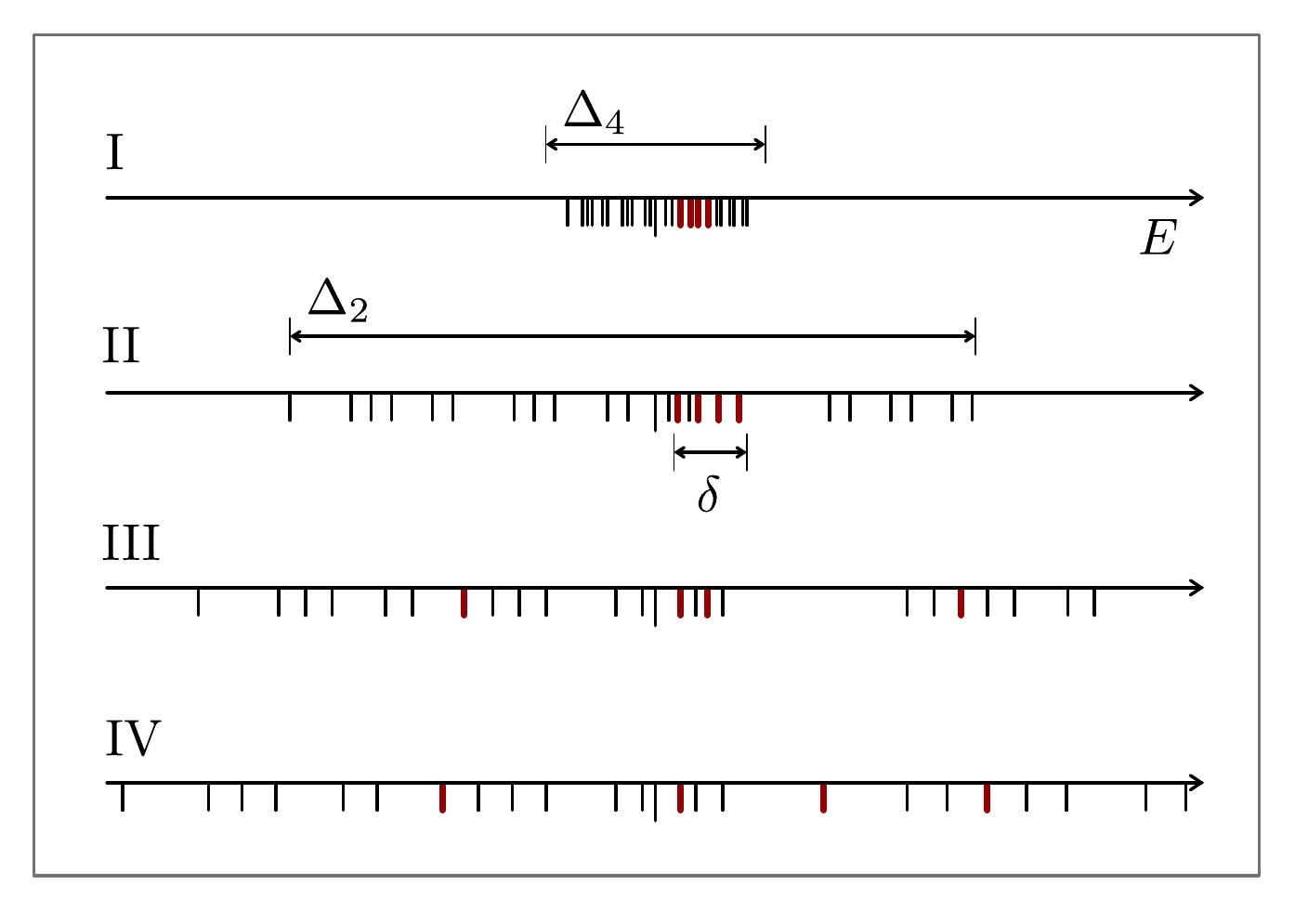}
\caption{\label{DisorderRegimes} 
The four regimes I-IV of increasing disorder strength. The figure indicates the band width $\Delta_4$ of the interaction operator in comparison to band width of the on-site randomness, $\Delta_2$. The distance between neighboring levels is $\sim \delta =\Delta_2 N^{-1/2}$, and the relative magnitude of these scales defines the regimes discussed in the text.}
\end{figure}

\begin{itemize}
    \item[I:] $\delta < \Delta_4 N^{-1/2}$: In this regime, the $\hat H_2$ band width $\Delta_2<\Delta_4$ is below that of $\hat H_4$. The on-site randomness is largely irrelevant and states are ergodically spread over the full Fock space lattice. Hybridization of levels over the full $\Delta_4$ band width implies $\kappa_n=\Delta_4$.
    \item[II:] $\Delta_4 N^{-1/2}<\delta< \Delta_4$: The $\hat H_2$ band width $\Delta_2$ exceeds $\Delta_4$ implying that the majority of sites become inaccessible.  States of  fixed energy now populate only a fraction of Hilbert space sites. However, for a given site with energy $v_n$ inside the accessible window $\Delta_4$, the hopping nearest neighbors have accessible energy $v_n \pm \mathcal{O}(\delta)$, and thus are also accessible. As a consequence, $\kappa_n=\Delta_4$ for all sites with energy $|v_n| \lesssim \Delta_4$.
    \item[III:] $\Delta_4<\delta<\Delta_4 N^{\alpha/2}$: In this regime, the energetic separation even between
    nearest neighbors $\delta > \Delta_4$ exceeds the interaction band width. In
    the consequence, the hybridization of levels with energy $v_n\approx 0$ is suppressed down to $\kappa_n\sim \Delta_4\times (\Delta_4/\delta)$, and the band of accessible
    sites narrows to this width. For a given site $n$ inside the resonant window,
    nearest neighbors of energy $\sim \mathcal{O}(\delta)$ typically lie outside it.
    However, a fraction $\sim (\Delta_4^2/\delta)/\delta=(\Delta_4/\delta)^2$ of the nearest neighbors does satisfy the resonance condition. With $\sim N^\alpha$ neighbors, this gives a number of $N^\alpha (\Delta_4/\delta)^2>1$ of hybridizing partner sites, which safeguards the extension of states. 
    \item[IV:] $N^{\alpha/2}\Delta_4<\delta$: The number of nearest neighbors satisfying the resonance condition becomes lesser than unity which implies strong localization of states in Fock space. 
\end{itemize}

The regimes I-IV cover the entire spectrum from fully extended states, I, over `non-ergodically' extended (NEE) states II,III, to localization IV. (In regimes II,III states cover only a fraction of the Fock space sites. In this paper, we are following the convention to call such non-uniformly distributed states `non-ergodic'. This is a misnomer in that the states do remain uniformly distributed over an `energy shell' of resonant sites.) The level broadening characterizing the local spectral density in the respective regimes is described by the universal formula
\begin{align}
      \label{KappaUniversal}
      \kappa_n \approx \kappa e^{- \frac{v_n^2}{\kappa^2}},
  \end{align}  
  where the value of the hybridization parameter and the corresponding disorder strengths are summarized in table~\ref{tab1}.

\begin{table*}
  \begin{tabular}{|l|l||c|c|c|c|}\hline
      & regime& disorder, $\delta$ & level broadening, $\kappa$ &spectral statistics&state extension \cr \hline
      I& RMT&$\delta N^{1/2}=\Delta_2 < \Delta_4$ &$\kappa\sim\Delta_4$ &Wigner-Dyson &$D$ \cr 
      II& NEE$_1$&$\Delta_4N^{-1/2}<\delta<\Delta_4$ &$\kappa\sim\Delta_4$&Wigner-Dyson &$ D \Delta_4/\sqrt{N}\delta$ \cr
      III & NEE$_2$&$\Delta_4<\delta<\Delta_4N^{\alpha/2}$ &$\kappa\sim \Delta_4^2/\delta$&Wigner-Dyson &$ D\Delta_4^2/\sqrt{N}\delta^2$\cr
      IV& localization&$\Delta_4 N^{\alpha/2}<\delta$ &$\kappa=0$ &Poisson &$\mathcal{O}(1)$\cr\hline
  \end{tabular}
  \caption{\label{tab1} Different regimes of disorder strength, the associated level hybridization, spectral statistics, and eigenfunction support in Fock space.
}
\end{table*}

Before leaving this section, it is worthwhile to comment on various phenomenological approaches to MBL. 
We distinguish between three categories of phenomenological formulations. The `most phenomenological' 
class models Fock space by a random matrix. For example, the Rosenzweig-Porter (RP) model contains a 
Gaussian distributed random matrix (as a proxy of the interaction operator, $\hat H_4$) 
perturbed by a likewise random diagonal representing $\hat H_2$~\cite{RPmodel,RPmodel1}. 
The second class replaces Fock space by a high dimensional synthetic lattice, 
such as the Bethe lattice~\cite{Zirnbauer86,NonE_Ex,Biroli18}, or a 
random regular graph~\cite{Mirlin1,TikhonovMirlin19_2,Lemarie17,Lemarie20}. 
Finally, there is the random energy model (REM) which retains the microscopic structure of 
Fock space  but replaces the  amplitudes $v_n$ by $2^N$ uncorrelated random variables 
(see our previous publication~\cite{NEE_SYK} for an application of this idea to the SYK Hamiltonian). 
These models are designed to mimic specific aspects of localization and wave function statistics 
in high dimensional environments. However, they  fall short of describing the characteristic correlations 
between site energies  and  high lattice coordination number essential to the distinction of the 
regimes I-IV and their statistical properties reviewed in the next section.   

One of the main messages of this paper is that the analytical theory for `real' systems 
need not be more difficult than that for synthetic models. What at first sight looks like a complication --- the 
combination of high coordination numbers and correlations in the microscopic Fock space  --- actually is a 
resource and leads to self averaging (a source of simplicity) at several stages of our computations below.

On this basis, we now discuss quantitative results obtained for the description of regimes I-IV. For notational simplicity, we  work in units where the variance of the $\hat H_4$ matrix elements equals $J= (2/N)^{1/2}$. At this value, the band width of the interaction
operator $\Delta_4\equiv \sqrt{J^2N/2}=1$.

\begin{figure}
\centering
\includegraphics[width=7cm]{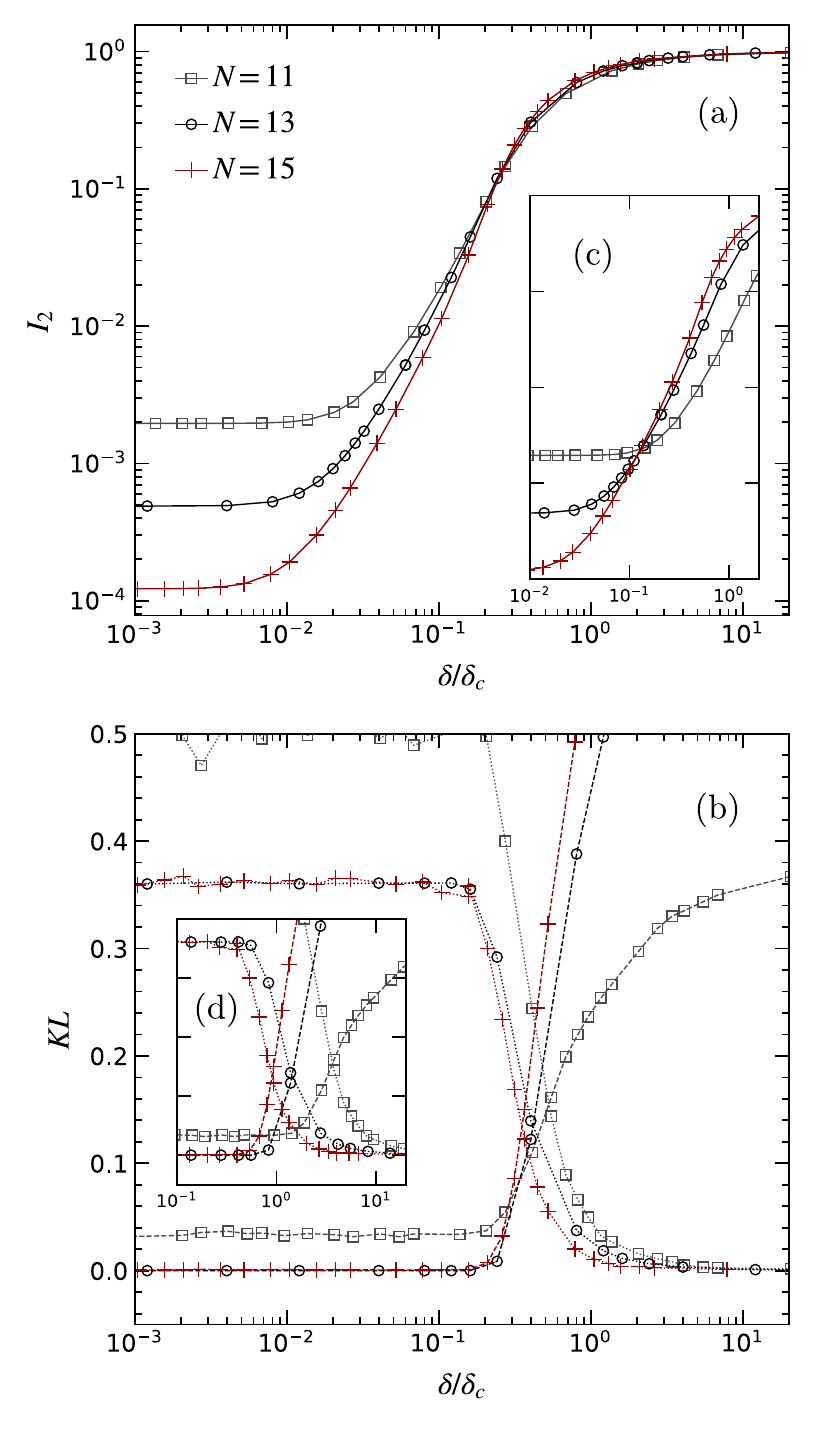}
\caption{\label{fig3} 
(a): Scaling of the inverse participation ratio $I_2$
 for system sizes
 $N=11, 13, 15$
as functions of the dimensionless disorder strength $\delta/\delta_c$, 
where $\delta_c$ is the critical strength 
 obtained by analytical solution of the model in Eq.~\eqref{app_dcLW}.
(b): plot of the relative Kullback-Leibler entropies ${\it KL}$
between the
numerical spectral statistics and the Wigner-Dyson (dashed) 
and Poisson distribution (dotted), respectively, for the same set of system sizes. 
In either case, the analytically obtained $\delta_c$ 
overestimates the critical strength by an  $N$-independent 
factor of ${\cal O}(1)$. 
(c), (d): Scaling of $I_2$ and ${\it KL}$
as functions of $\delta/\delta_c$, 
employing Eq.~\eqref{deta_c_d} with two adjusted numerical parameters, see discussion in 
main text.  
}
\end{figure}

\subsection{Spectral statistics} 
\label{sub:res_spectral_statistics}

We describe the statistics of the system's many body spectrum in terms
of the spectral two point correlation function at the band center
\begin{align}
\label{K}
&K(\omega)
\equiv
{1\over \nu^2}
\langle \nu(
\tfrac{\omega}{2}) 
\nu(
-\tfrac{\omega}{2}) \rangle_{c},
\end{align}
where
$\nu = \nu(E= 0)$ with
$\nu(E) = \sum_\psi \langle\delta(E-\epsilon_\psi)\rangle_J$
is the $\hat H_4$ averaged many body density of states
at zero energy $E\simeq0$ and
the subscript $c$ stands for the cumulative average
$\langle AB\rangle_c=\langle AB\rangle_{J} - \langle A \rangle_{J}\langle B\rangle_{J}$.

\emph{Regimes I-III:} In these regimes, wave functions are extended and their eigenenergies are correlated 
and described by Wigner-Dyson statistics. 
 Assuming an odd number $N$ of complex fermions (for which the SYK model is in the unitary symmetry class A), this reflects in the spectral statistics of the Gaussian unitary ensemble (GUE),
\begin{align}
     \label{GUESpectralStatistics}
     K(s)=1-\frac{\sin^2s}{s^2}+\delta(s/\pi), \qquad s=\pi \omega \nu,
 \end{align} 
 where $ \nu=\sum_n \nu_n$ is the average density of states. With the local densities given by Eq.~\eqref{LocalDosLorentzian}, and the $v_n$ distributed over a range $N^{1/2}\delta$, we find 
 \begin{align}
     \label{DoSAverage}
     \nu\equiv \sum_n \nu_n=cD\left\{
\begin{array}{cl}
    1&\mathrm{I},\cr
    \frac{1}{\sqrt N \delta}&\mathrm{II, III},
\end{array}
     \right.
 \end{align}
 where, here and throughout, $c=\mathcal{O}(1)$ represents numerical constants. The second line of Eq.~\eqref{DoSAverage} states that in the regimes of
intermediate disorder strength, only a fraction $D/\sqrt{N}\delta$ of active sites
contributes to the spectral support of wave functions. 

\emph{Regime IV:} In the regime of strongly localized states, eigenenergies become uncorrelated and we expect Poisson statistics. In this paper, we will use the change from Wigner-Dyson to Poisson statistics as one of two indicators for the Anderson transition at the boundary between regimes III and IV. Referring for a more detailed discussion of the localization transition to Section~\ref{sub:strong_localization} below, we note that in the literature~\cite{Huse10}, the difference between the
two types of statistics is often monitored by analysis of $r$-ratios~\cite{Huse07},
i.e.\ numerical comparison of the ratios
$r_k \equiv 
\frac{\epsilon_{k+1}-\epsilon_k}{\epsilon_{k}-\epsilon_{k-1}}$ between nearest neighbor many body energy levels,
$\epsilon_k$, 
with the  expected ratios for Poisson and Wigner-Dyson statistics. However, we
have observed that naked eye comparisons can easily trick one into premature and
qualitatively wrong conclusions. Instead, we adopt a more sophisticated entropic
procedure detailed in Sec.~\ref{sub:comparison_to_numerics} and compute 
Kullback-Leibler divergences, where  the latter are defined as relative entropies of the numerically observed
distribution to the Poisson and Wigner-Dyson distribution, respectively. The panel (b) of Fig.~\ref{fig3} shows how this entropic measure changes abruptly at the localization transition.

\subsection{Wave function statistics} 
\label{sub:res_wave_function_statistics}

The second class of observables considered in this paper are the
moments of wave functions $|\psi\rangle$ of zero energy, $\epsilon_\psi = 0$,
 \begin{align}
\label{Iq}
&I_q \equiv
{1\over \nu}
\sum_n 
\langle 
|\langle \psi|n\rangle |^{2q} \,\delta(\epsilon_\psi)
\rangle_{J}.
\end{align}
The statistics of these moments not only indicates the localization transition but, unlike spectral statistics, also differentiates between the three weak disorder regimes I-III.

\begin{figure*}
\centering
\includegraphics[width=16.5cm]{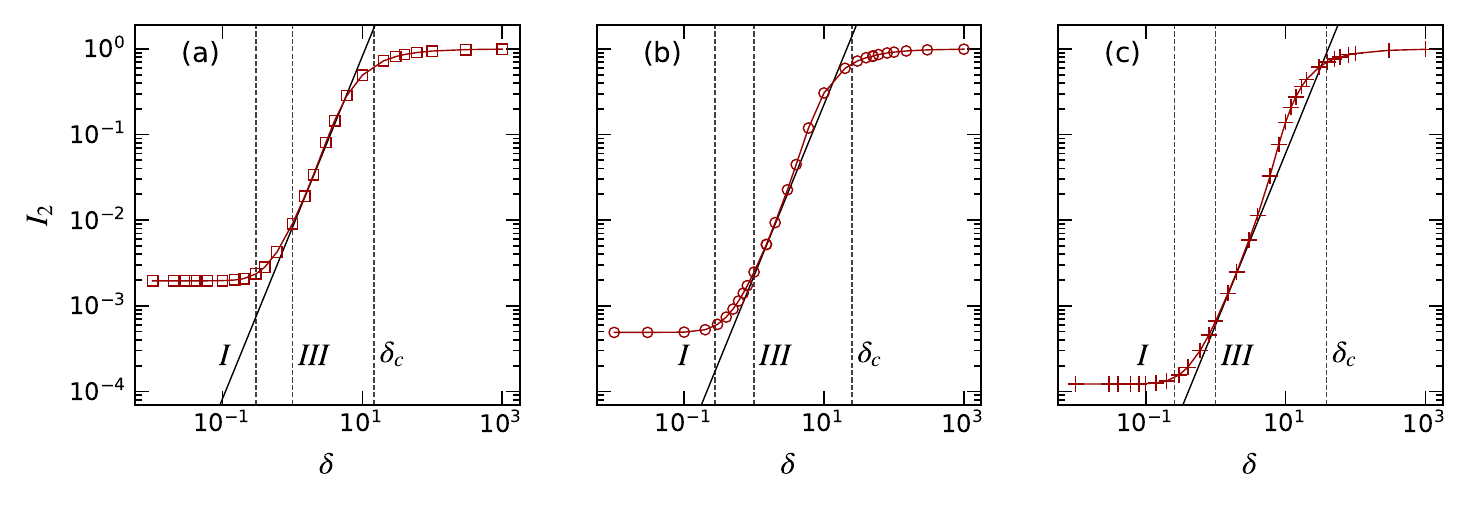}
\vspace{-.4cm}
\caption{\label{fig1} 
Comparison of numerical computation of inverse participation ratio $I_2$ as a function of the 
disorder strength $\delta$ 
for system sizes $N=11$ (a), $13$ (b), and $15$ (c) with the analytical prediction
$I_{2} = 8\sqrt{N} \delta^{2}/(\pi D)$, see Eq.~\eqref{app_I_q_sc}. 
Vertical dashed lines mark the end of region ${\rm I}$, beginning of region ${\rm III}$, and the
scale $\delta_c$ at which Fock space localization sets in 
(estimated from Eq.~\eqref{app_dcLW})~\cite{fn8}. 
}
\end{figure*}

\emph{Regime I:} wave
functions are ergodically distributed over the full Fock space, with moments given by those of the Porter-Thomas distribution, 
\begin{align}
    \label{MomentsI}
    I_q
&=q! D^{1-q},\qquad \text{I},
\end{align}
otherwise found for the wave functions of random matrix Hamiltonians. The result
states that the complex amplitudes $\langle n|\psi\rangle$ are independently distributed Gaussian random variables.

\emph{Regimes II \& III:} The wave functions no longer ergodically occupy the
full Fock space. The bulk of their support is concentrated on the subset of resonant sites, $v_n \sim \kappa_n$. This behavior reflects in the moments
\begin{align}
    \label{MomentsIIAndIII}
    I_q = c^{q}\left(\frac{D}{\sqrt{N}\delta}\right)^{1-q}\frac{2q(2q-3)!!}{\kappa^{q-1}},\qquad \text{II, III},
\end{align}
where $c=\mathcal{O}(1)$. To make the connection of this expression to Eq.~\eqref{MomentsI} more transparent, consider the case of large $q$, where
\begin{align}
    \label{IqRegimeII-III}
    q\gg 1: &\qquad I_q=cq! D_\mathrm{res}^{1-q},\cr
    &D_\mathrm{res}=D\left\{
\begin{array}{ll}
    \frac{1}{\sqrt{N}\delta},& \text{II},\cr
    \frac{1}{\sqrt{N}\delta^2},& \text{III}.
\end{array}\right.
\end{align}
These moments again coincide with those of a Gaussian
distribution, now defined on the diminished number $D_\mathrm{res}$ of resonant sites in
Fock space, over which the wave functions are uniformly spread. 

Noting that $\delta \sim N^\eta$, $\eta<2$,
the dependence of $D_{\mathrm{res}}$ on $D$ is approximated as
\begin{align}
  D_\mathrm{res} = D/\log D^{\beta}, \quad \beta = \begin{cases}
    \eta+1/2  &\text{II},\\
    2\eta+1/2 &\text{III}.
    \end{cases}
\end{align}
 
This suggests an interpretation in terms of a `fractal'  
whose dimension differs from the naive dimension by a factor $D/\log D^\beta\sim D/D^0$, rather than the more usual
$D/D^\gamma$ with some $\gamma>0$.  
Alternatively, we may interpret the wave functions as ergodically, or \emph{thermally} extended over an `energy shell' of sites defined by the condition $v_n \approx \kappa_n$.

\begin{figure}
\centering
\includegraphics[width=7cm]{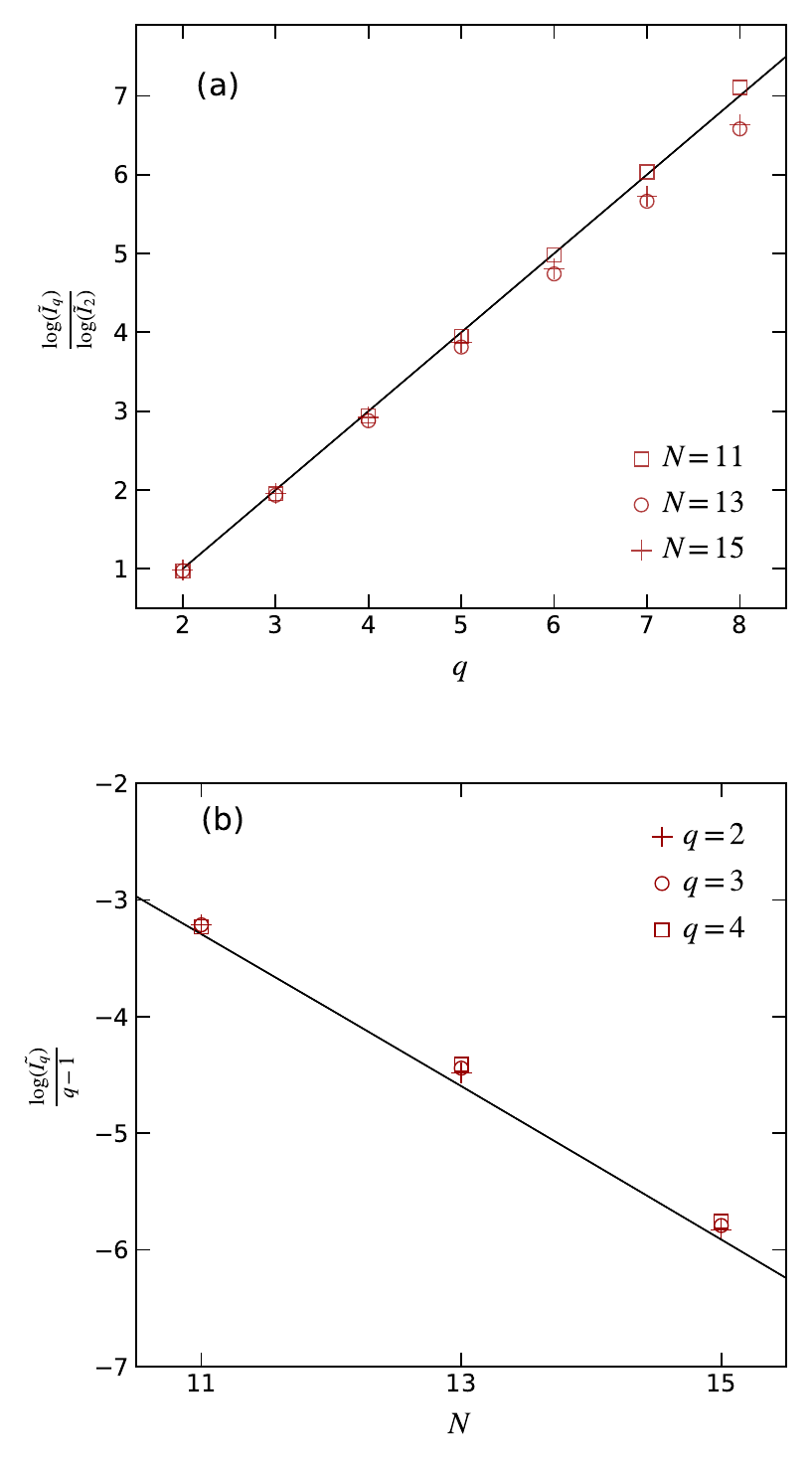}
\caption{\label{fig2} 
Verification of the scaling of our analytical prediction Eq.~\eqref{MomentsIIAndIII} 
in $q$ and $N$, respectively. In both panels we consider $\delta = 3$ deep in regime III and $\tilde{I}_q\equiv I_q/[q(2q-3)!!]=(4\sqrt{N} \delta^2/\pi D)^{q-1}$, where the constants are taken from the accurate result for $I_q$ in regime III, Eq.~\eqref{app_I_q_sc}. 
}
\end{figure}

Figures~\ref{fig1} and~\ref{fig2} show a comparison of our analytical
predictions for the wave function moments dependence on 
$\delta$ (Fig.~\ref{fig1}), respectively $q$ and $N$ (Fig.~\ref{fig2}), 
to numerical simulations 
for $2N=22, 26, 30$ Majorana fermions.
Vertical dashed lines in 
 Fig.~\ref{fig1} mark the boundaries between different regimes, and $\delta_c$ is the scale at which Fock space localization sets in, see 
 Eq.~\eqref{deta_c_d} and the refined expression, Eq.~\eqref{app_dcLW}, accounting for $1/N$ corrections. 
 For the numerically accessible $N$-values, regime II, $N^{-1/2}\ll \delta\ll
 1$, lacks the width required for the comparison with power laws and we concentrate on regime III. Given that there is no fitting of numerical parameters and numerical error bars are smaller than symbol size, the comparison is good. We notice a slight deviation in the $q$-scaling, increasing for large moments, $q$. However, this mismatch does not show consistent system size dependence, and we cannot attribute a clear trend to it.
 Starting from $N=13$ we also see deviations of the predicted $\delta$-scaling at large values, which is a first indication of the proximity of the Anderson transition. At first sight, it may seem paradoxical that these signatures are first seen for larger $N$, where the parametric dependence of the localization threshold $\delta \gtrsim N^2$ increases in $N$. However, the situation becomes clearer when we represent the inverse participation data as a function of a scaled parameter, as we will discuss next. 
 
 \subsection{Strong localization} 
\label{sub:strong_localization}

The wave functions describing random hopping on a lattice are localized on small sized
clusters if, statistically, the nearest neighbor hopping matrix elements become
smaller than the  variations of the local site energies. In this work, we numerically and analytically compute the threshold strength of the disorder where this happens. 

\begin{figure}
\centering
\includegraphics[width=6.5cm]{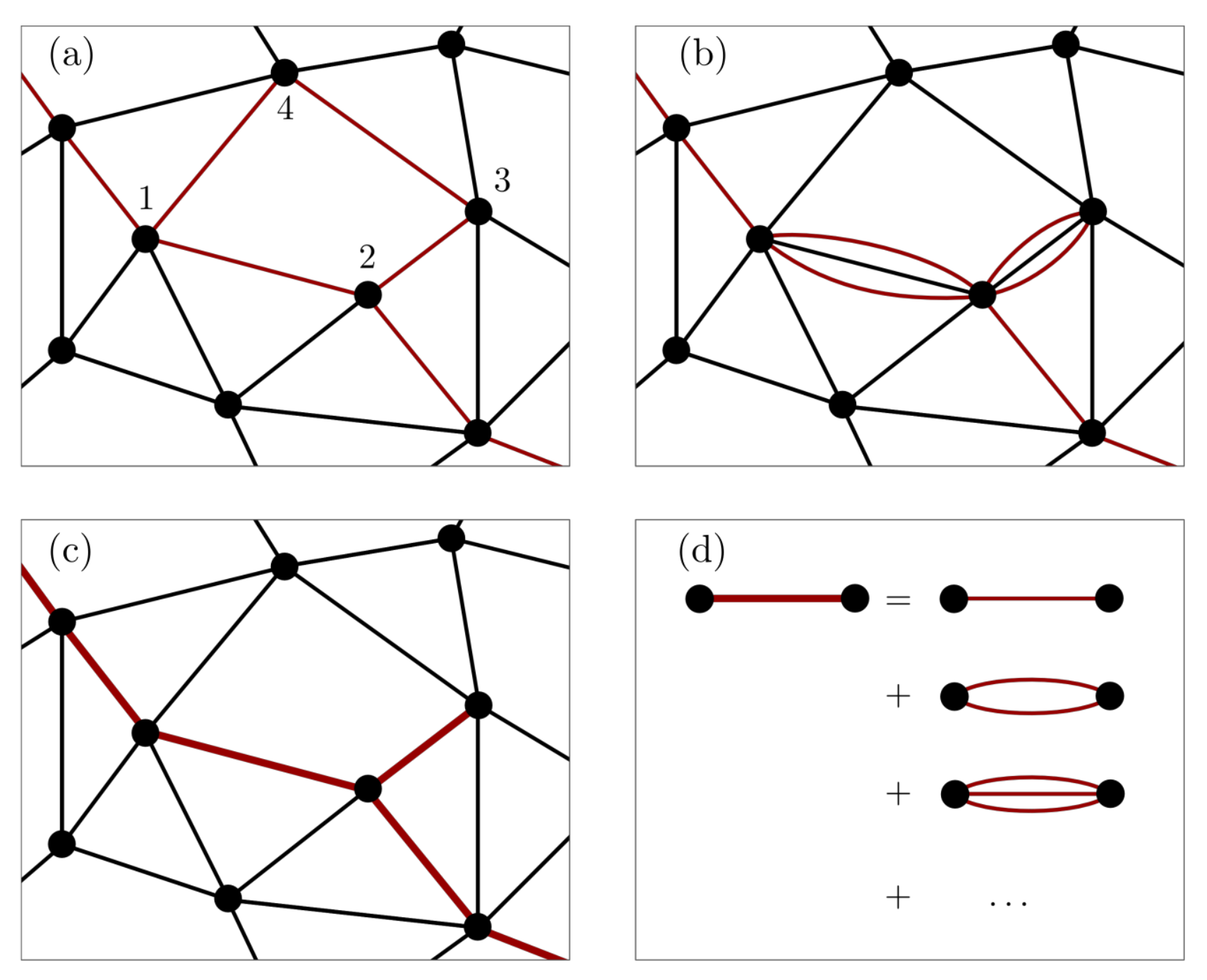}
\caption{\label{fig:NoLoop} 
(a): a cartoon representation of a subset of sites in Fock space connected by a hooping amplitude containing a loop insertion. The four hopping amplitudes constituting the loop come with four independent energy denominators. (b): this fourth order hopping amplitude with site revisits has only two independent energy denominators and contributes parametrically stronger. (c): hopping amplitudes resumed according to the procedure shown in the (d).
}
\end{figure}

\emph{Analytical approach:}  Above, we reasoned that the problem of MBL is defined by
a competition of localizing on-site disorder and delocalizing hopping in a complex
high-dimensional lattice. Unlike in previous  work on Anderson localization in high
dimensions, which is formulated on simplified synthetic lattice structures such as
the Bethe lattices~\cite{Kravtsov2,Biroli17} or  random regular
graphs~\cite{Kravtsov2,Biroli12,Kravtsov1,DeLuca14}, we here directly work in Fock
space. What helps to keep this more complicated problem under control is the huge
effective lattice coordination number of $\mathcal{O}(N^4)$, and a simplification
known as the `effective medium approximation'~\cite{Efetov}. This approximation is
commonly applied in the discussion of Anderson localization on high dimensional
lattices and backed by their large coordination numbers. It describes transport as a
process avoiding local loops (see Fig.~\ref{fig:NoLoop}), while multiple link
traversals (panel (a)) are included. The rationale behind this simplification is that
at any given order in hopping perturbation theory, amplitudes with the lowest number
of statistically independent energy denominators contribute the strongest. Its
application to the SYK lattice, detailed in Section~\ref{Ergodic_loc}, sums these
processes via recursion relations (such as Eq.~\eqref{eq:EffectiveMediumLinearized})
whose solution leads to prediction Eq.~\eqref{app_dcLW} for the critical disorder
strength.
For large $N\gg 1$ this formula simplifies to
\begin{align}
\label{deta_c_d}
\delta_c
&\simeq
{N^2\over 4\sqrt{3}}
\log N, \qquad (N\gg 1).  
\end{align} 
The  characteristic $\delta_c\sim N^2\log N$ scaling was first predicted in Ref.~\cite{AltshulerGefenKamenevLevitov97}, 
where the logarithmic correction relative to the naive estimate
$\delta_c \sim N^2$ mentioned in the introduction accounts for resonant hybridization
with sites beyond nearest neighbors. 

However, our aim here is to compare to the real world of  small sized systems
$N=\mathcal{O}(10)$ where things get more complicated. For one, the difference
between the asymptotic result and the more precise expression Eq.~\eqref{app_dcLW}
becomes noticeable. Second, various approximations in the execution of the effective
medium program rely on the largeness of $N$ and again may lead to errors in terms
sub-leading in $N$. These uncertainties must be kept in mind when we compare to the
numerical computation of the threshold.

\emph{Numerical approach: } As indicated above,  we detect the onset of localization via two indicators. The first is
the wave function statistics, where $I_2$ serves as a transition order parameter
jumping between the values $I_2 \sim D^{-1}$ in the ergodic weak disorder regime to
$I_2\sim 1$ in the localized phase. Here, the first value must be taken with a grain
of salt, again due to finite system size. Our discussion in the previous section shows that before reaching the transition, in regimes
II, III, we have deviations away from the ergodic limit $I_2\sim 1/D$. In the
thermodynamic limit, these are inessential (because $D$ is exponential in $N$ while
the corrections of Eq.~\eqref{IqRegimeII-III} are in powers of $N$.) However, for
system sizes in numerical reach, we cannot expect an actual jump in the order
parameter. The best one can hope for is gradual steepening of the curve $I_2(\delta)$
for $\delta\to
\delta_c$ upon increasing system size. 

The second diagnostic is spectral statistics, where we monitor the proximity to a Wigner-Dyson or Poisson distribution via the Kullback-Leibler entropy as discussed in Section~\ref{sub:spectral_statistics}.
Ideally, one would hope that both signatures, inverse participation ratio and
spectral statistics reveal a phase transition via a crossing point when subjected to
appropriate finite size scaling, and that these crossing points sit at the same
value. In reality, we almost, but not fully observe this behavior. In
Fig.~\ref{fig3}, we show the inverse participation ratio $I_2$, and the
Kullback-Leibler entropy as a function of the scaled variable, $\delta/\delta_c$,
where $\delta_c$ is given by the analytical prediction Eq.~\eqref{app_dcLW} in terms of the Lambert
$W$-function. We observe that (i) both observables show reasonably well defined
crossings with a tendency of sharpening behavior for increasing system size, however,
(ii) these crossings deviate from the analytically predicted value
$\delta/\delta_c=1$ by a numerical factor of $\mathcal{O}(1)$, and by a factor of
similar magnitude among themselves. Turning to different scaling variables, one may sharpen the finite size scaling of either one of the two observables. 
For example, the inset in the first panel shows $I_2$ as a function of $\delta/\delta_c$, with $\delta_c$ from 
  Eq.~\eqref{deta_c_d} with two numerical parameters outside and inside the `log'
  adjusted  to improve visibility of the crossing point~\cite{crossing}. However, this comes at the expense of a more diffuse scaling of the
 entropy, as shown in the inset of the second panel. We observe that the numerically obtained scaling for small systems responds sensitively to the
 finite $N$ corrections (Eq.~\eqref{app_dcLW} vs. Eq.~\eqref{deta_c_d}). 

All in all, we consider the agreement with the numerics quite favorable. We see clear evidence of critical behavior in two observables and the position of the transition is obtained without free fit parameters from the analytical solution of an effective  lattice model. This may be the first genuine Fock space localization problem where a first principle solution of this kind has been possible. 

In the next sections, we discuss the derivation of the analytical results mentioned above. Hoping that elements of this computation might become blueprints for the analysis of other models of MBL, we try to be as pedagogical as we can. Various technical details are relegated to appendices.


\section{Matrix model}
\label{M_model}
We start the derivation of the results summarized above by constructing an exact
matrix integral representation of the correlation functions introduced above to
describe many body wave functions and spectra. The unconventional perspective of this
approach is that there will be no second quantized representation of Fock space: we
think of the SYK Hamiltonian as a big matrix, and treat it like that. 
In this section, we discuss the construction of a matrix integral representing the theory averaged over $\hat H_4$ disorder. The physics behind
this formulation and that of a subsequent stationary phase analysis of the theory will be
discussed in the next section.

All information on spectra and wave functions of the system is contained in the Fock space matrix elements of resolvent operators,
\begin{align}
G^{\pm}_{nm}
&
=
\big\langle n | 
(z_\pm  
-\hat H)^{-1}|m\big
\rangle, 
\end{align}
where $z_\pm=\pm(\frac{\omega}{2}+i\eta)$ and, here and throughout, 
$\eta$ is infinitesimal (with a limit $\eta\searrow 0$ to be taken in the final
step of all calculations). Specifically, the correlation functions above are obtained as
\begin{align}
\label{IqGtreenFunctions}
&I_q
=
\frac{(2i\eta)^{q-1}}{2i\pi \nu} \sum_n
\langle 
G^{+(q-1)}_{nn}G^-_{nn}
\rangle_{J},
\nonumber \\
&K(\omega)
=
{1\over 2\pi^2\nu^2} 
\sum_{nm} 
{\rm Re} \langle 
G^+_{nn}
G^-_{mm}
\rangle_{J},
\end{align}
where $I_q$ is computed at $\omega=0$, and $\langle\cdots\rangle_{J}$ denotes the average over
coupling constants $\{J_{ijkl}\}$ of $\hat H_4$.

\emph{Construction of the matrix integral. ---} Following standard protocols, we raise the Green functions to an exponential representation before performing the Gaussian average. The basic auxiliary formula in this context is 
 $M^{-1}_{nm}=
  \int D(\bar\psi,\psi)\,e^{-\bar \psi M\psi}\psi^\sigma_m\bar \psi^\sigma_n$, 
where $M$ is a general $L\times L$ matrix and the $2L$ dimensional `graded' vector $\psi=(\psi^\mathrm{b},\psi^\mathrm{f})^T$
contains $L$-commuting components $\psi_n^\mathrm{b}$, and an equal number of Grassmann 
components $\psi_n^\mathrm{f}$.
The double integral over these variables cancels unwanted determinants 
$\det(M)$, while the pre-exponential factors, either commuting or anti-commuting, $\sigma=\mathrm{b,f}$, isolate the inverse matrix element. With the identification
 $M=\mathrm{diag}(-i [G^+]^{-1},i [G^-]^{-1})
 =
 -i\sigma_3( 
 E + z  -\hat H)$, 
we are led to consider the generating function
\begin{align}
\label{eq:Zpsi}
    {\cal Z}[j]
=
\int D(\bar{\psi},\psi)
\left\langle e^{-\bar{\psi}
\left(
E + z -\hat H
- j
\right)\psi}\right\rangle_{J}.
\end{align}
Here, 
$z \equiv \left( \tfrac{\omega}{2} +i\eta \right)
\sigma_3$, contains the energy arguments of the Green functions and $\sigma_3$ is a Pauli matrix distinguishing between advanced and retarded components. The matrix $j$ acts as a source for the generation of the required moments of Green function matrix elements. Specifically, we define
  \begin{align}
  \label{sourceK}
 j_K(\alpha,\beta)
&=
\alpha \pi^{\rm b}\otimes\pi^{\rm +} 
+
\beta \pi^{\rm f}\otimes\pi^{\rm -}, 
  \\
  \label{sourceI}
  j_{I,n}(\alpha,\beta)
&= 
 j_K(\alpha,\beta) \otimes |n\rangle\langle n|,
\end{align}
where $\pi^{\mathrm{b,f}}$ is a projector onto commuting and anticommuting-variables, respectively, $\bar \psi \pi^{\sigma}\psi=\bar \psi^\sigma \psi^\sigma$, and $\pi^{\pm}$ projects in causal space, $\bar \psi \pi^{s}\psi=\bar \psi^s \psi^s$, $s=\pm$. With these definitions, an elementary computation shows that  
\begin{align}
\label{gfK}
&K(\omega)
={1\over 2\pi^2\nu^2}  {\rm Re}\,
\partial^2_{\beta\alpha}{\cal Z}[j_K]|_{\alpha,\beta=0},
\\
\label{gfIq}
&
I_{q}=
c_q
(2i\eta)^{q-1} 
\sum_n\partial_\beta\partial_\alpha^{q-1} {\cal Z}[j_{I,n}]
|_{\alpha,\beta=0},
\end{align}
with 
 $c_q\equiv 1/(2i\pi \nu(q-1)!)$. In the following, we consider the sources absorbed in a redefined energy matrix, $z \rightarrow z-j$, and remember their presence only when needed. 

 At this point, the averaging over $\hat H_4$ can be performed, and it generates a quartic term 
 \begin{align}
    \label{SPsiQuartic}
    {\cal Z}
=
\int D(\bar{\psi},\psi)
 e^{-\bar{\psi}
\hat G^{-1}
\psi+\frac{w^2}{2}\sum_a (\bar \psi \hat X_a \psi)^2},
 \end{align} 
where we defined $w^2=6J^2/(2N)^3\equiv\frac{3}{2}N^{-4}$ for the scaled variance of the SYK Hamiltonian $\hat H_{4}$, $\hat G\equiv (E+z - \hat H_2)^{-1}$,
\begin{align}
    \label{XaDef}
    \hat X_a\equiv
\hat{\chi}_i\hat{\chi}_j\hat{\chi}_k\hat{\chi}_l,
\end{align}
and $a=(i,j,k,l)$ with $i<j<k<l$. We next perform an innocuous but physically
meaningful (see next section) rearrangement $(\bar \psi \hat X_a
\psi)^2=\STr((\psi\bar \psi \hat X_a)^2)$, where the supertrace~\cite{Efetov}
$\mathrm{STr}(X)\equiv \tr(X^{\mathrm{bb}})-\tr(X^{\mathrm{ff}})$ accounts for the
minus sign caught when exchanging anti-commuting variables. The next step is a
Hubbard-Stratonovich transformation decoupling the matrices $\psi\bar \psi \hat
X_a\sim  A_a$ in terms of $(2N)^4/4!$ auxiliary matrix fields $ A_a$.
Referring for details of the procedure to
Appendix~\ref{sec:derivation_of_the_action} we note that after the decoupling the
 integral over $\psi$-variables has become Gaussian and can be carried out. A more interesting statement is that of the $\rho\equiv \binom{2N}{4}$ Hubbard-Stratonovich fields $ A_a$, all but one can be removed, too, by straightforward Gaussian integration. Upon restricting to $E=0$ 
this leaves us with a single integration, 
\begin{align}
 \label{YAction}
   &{\cal Z}=\int {\cal D}Ye^{- S[Y]},\\
   &S[Y]
=
-{1\over 2}
{\rm STr}( Y \mathcal{P} Y )
+
{\rm STr}\log\left(
z 
-\hat H_2 
+
i \mathcal{P}Y
\right),\nonumber
\end{align}
over a $2\times 2 \times D$ dimensional
matrix $Y=\{Y_{nn'}^{\sigma\sigma',ss'}\}$ 
carrying indices in causal space, super-space, and Fock space. The
information on the SYK system now sits in the site-diagonal one-body term, $\hat
H_2$, and the hopping operator ${\cal P}$ which represents the interaction and acts on
matrices $Z=\{Z_{nm}\}$ in Fock space as
\begin{align}
      \label{PDef}
      \mathcal{P}Z\equiv \frac{1}{\rho}\sum_a \hat X_a Z   \hat X^\dagger_a. 
  \end{align}  
 Finally, $\gamma=w\rho^{1/2}= 1$ represents the $\hat H_4$ band width, which we have set to unity.
 To simplify formulas, we will consider energies 
 $\hat H_2 \to \gamma \hat H_2$, $\omega\to \gamma \omega$ scaled by this parameter, and suppress it throughout.

\begin{figure}
\centering
\includegraphics[width=5.5cm]{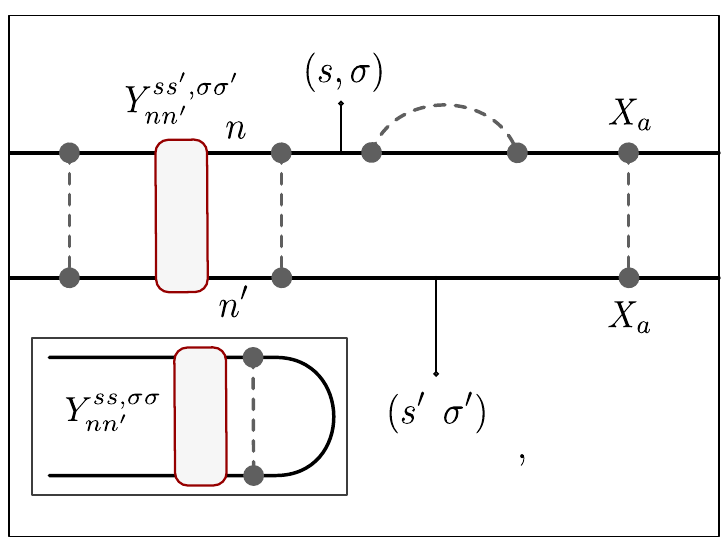}
\caption{\label{YVertex} 
The composite matrix degree of freedom $Y_{nn'}^{ss',\sigma\sigma'}$ representing the pair propagation of Fock space scattering amplitudes. Discussion, see text. 
}
\end{figure}

\emph{Discussion of the matrix integral. ---} 
This is now a good point to discuss the meaning of the above Hubbard-Stratonovich
transformation and of the matrix-representation. The two-fermion vertices $\bar \psi
\hat X_a \psi$ entering the theory after disorder averaging describe the scattering
of Fock space states off the four-Majorana operators contained in the Hamiltonian,
and in this way introduce the lattice connectivity indicated in Fig.~\ref{FockSpace}.
While a direct analysis of individual Fock space amplitudes seems hopeless, progress
can be made if the propagators are paired to two-amplitudes composites as indicated
in Fig.~\ref{YVertex}. For two reasons, the pair amplitudes
$Y^{ss',\sigma\sigma'}_{nn'}=\psi^{s\sigma}_n\bar \psi^{s'\sigma'}_{n'}$ are more
convenient degrees of freedom: First, the pair action $Y \to \sum_a \hat X_a Y \hat
X_a=\rho \mathcal{P}Y $ governing scattering in the two-state channel (cf.\ the
structure of the action~\eqref{YAction}) is relatively easy to describe, see below.
Second, the advanced/retarded combinations
$Y^{-+,\sigma\sigma'}_{nn}=\psi^{-\sigma}_n\bar \psi^{+\sigma'}_{n}$ appear as
terminal vertices in the computation of Green functions $G^+_{\cdot  n}G^-_{n\cdot}$,
where the dots stand for the unspecified final points of the correlation function.
With the exact identity $(G^+)^{-1}- (G^-)^{-1}=\omega^+\equiv \omega + 2i0$, we have
$\langle G^+_{mn}G^-_{nm}\rangle_{J} = \langle \tr(G^+ G^-)\rangle_{J}
= \frac{1}{\omega^+}\langle \tr(G^{+}) [(G^+)^{-1}- (G^-)^{-1}] G^-\rangle_{J} =
\frac{1}{\omega^+}\langle \tr(G^{-} -G^+)\rangle_{J} \simeq \frac{2\pi i}{\omega^+}\nu$,
where $\nu$ is the density of states at the band center. The way to read this (Ward)
identity is that the product of Green functions contains a singularity,
\emph{provided} $\tr(G^--G^+)\sim \nu$ is a structureless quantity. (The latter
condition does not hold in systems with localization, where the isolated eigenstates
support a point spectrum with poles rather than a uniform cut.) This argument
indicates that the `soft mode' $G^+G^-\sim \omega^{-1}$ is key to the understanding
of observables probing spectrum and eigenfunctions of the system.

In the matrix integral framework, the above singularity shows in the presence of a soft mode in the integration over the variables
$Y^{-+,\sigma\sigma'}_{nn}$. 
To isolate this mode, we note that Eq.~\eqref{YAction} has an approximate symmetry 
\begin{align}
    \label{ZereModeSymmetry}
    Y \to T Y T^{-1},\qquad T=\{T^{ss',\sigma\sigma'}\}
\end{align}
under rotations homogeneous in Fock space. The set of these transformations defines
$\mathrm{GL}(2|2)$, i.e.\ the group of invertible $4\times 4$ matrices with
anti-commuting entries. Invariance under this symmetry is weakly broken only by the frequency/source matrix $z$, which, ignoring the infinitesimal sources, transforms as
$\frac{\omega}{2} \sigma_3\to \frac{\omega}{2}T^{-1}\sigma_3 T$. This reduces the
symmetry down to the transformations diagonal in advanced-retarded ($s$-indices)
space, $\mathrm{GL}(1|1)\times \mathrm{GL}(1|1)$.

The essential question now is whether the above weak explicit symmetry breaking is
\emph{spontaneously broken} in the matrix integral (much as a weak explicit symmetry
breaking by a finite magnetic field gets upgraded to spontaneous symmetry breaking in
a ferromagnetic phase.) In the latter case, we expect a soft Goldstone mode whose
`mass' is set by the symmetry breaking parameter, $\omega$, and $\omega^{-1}$
singularities in line with the observation above. To investigate this question and
the consequences in the observables $K(\omega)$ and $I_q$, we next subject the theory
to a stationary phase analysis.

\section{Effective theory}
\label{Effective_T}

In this section, we map the exact theory Eq.~\eqref{YAction} to an approximate but
more manageable effective theory. The conceptual steps are standard and consist of
 a saddle point analysis, followed by a Ginzburg-Landau style expansion (see Section~\ref{Ergodic_loc}) 
 of the exact action 
  in fluctuations around a homogeneous saddle point.

We have already established the presence of an
exact (in the limit $\omega\to 0$) rotational soft mode isotropic in Fock space.
Since much of the analysis below will focus on strong $\hat H_2$ with eigenvalues
$v_n$ of $\hat H_2$ comparable to or exceeding the $\hat H_4$, we anticipate that
fluctuations of lowest action cost will be commutative in the sense $[\hat H_2,Y]=0$.
We thus start from an ansatz $\langle n|Y|m\rangle=Y_n\delta_{nm}$ where fluctuations
are diagonal in the occupation basis. In view of the fermion parity conservation of
both $\hat H_{2,4}$ we focus on a sector of definite parity, chosen to be even. The
locality of $\hat H_2$ in the occupation number basis is in competition with the
hopping described by $\mathcal{P}$. However, what works to our advantage is that the
action of $\mathcal{P}$ on the states $Y$ is remarkably simple: thinking of $Y_{nm}$
as the matrix elements of a `density matrix', $Y_n$ represents a state without
off-diagonal matrix elements. It is a non-trivial feature of $\mathcal{P}$ that it
preserves this structure, $(\mathcal P Y)_m=\sum_n \mathcal P_{|n-m|}Y_m$, i.e.\ the
adjoint action $\hat X_a Y \hat X_a$ on Fock-space diagonal matrices $Y$ does not
generate superpositions of off-diagonal states. A straightforward combinatorial
exercise shows that (see Appendix~\ref{app_operator_P} for details)
\begin{align}
\label{eq:PMatrixElements}
\mathcal{P}_0
=\frac{N(N-1)}{2\rho}, \quad \mathcal{P}_2=\frac{4(N-2)}{\rho},\quad \mathcal{P}_4=\frac{16}{\rho},
\end{align}
with all other matrix elements vanishing, and normalization $\sum_m
\mathcal{P}_{m,n}=1$. Notice that for a given $n$, we have $\binom{N}{4}$ neighbors
with hamming distance $4$, connected to $n$ by $\binom{N}{4}\mathcal{P}_4 \stackrel{N\gg
1}{\sim} 1$. This shows distance $4$ hopping is the most important 
by `phase volume'.

With these structures in place, a 
 variation of action~\eqref{YAction} leads to 
 \begin{align}
\label{mf}
   -iY 
   &= 
    \frac{1}{z -  \hat H_2 +i  \mathcal{P}Y  }.
\end{align}
Notice that $Y$ resembles $(-i)$ times the local `propagator' (see inset of Fig.~\ref{YVertex}) of site $n$, dressed with a self energy $i \mathcal{P}Y$ due to hopping via $\mathcal{P}$ to neighboring sites. It is this term which makes the stationary phase equation non-trivial. In a first step towards the solution, we neglect imaginary contributions to $Y$ and focus on the local `spectral density' $\mathrm{Re}(Y)$ instead. (In the effective action, the imaginary part of $Y$ describes an energy shift $v_n \to v_n + \mathrm{Im}Y_n$ which is inessential to our problem.) Causality requires $\mathrm{sgn}\,Y=\mathrm{sgn\;Im\,}z$, i.e.\ the sign of
the self energy is dictated by that of the imaginary part contained in the energy
arguments. Otherwise the saddle point equation is rotationally invariant in the
internal indices of the theory. This motivates an ansatz,
\begin{align}
   \mathrm{Re}\,Y=\sum_n
(\mathrm{Re}Y)_n|n\rangle\langle n|
\equiv \sum_n \pi \nu_n|n\rangle\langle n| \otimes\sigma_3\otimes\openone_{\rm bf} 
\end{align}
with real coefficients $\nu_n$. Inspection of Eq.~\eqref{mf} shows that these coefficients afford an interpretation as mean field local density of states. 

 Substituting this expression into the equation and
temporarily ignoring the small energy argument, $z$, as small compared to both
$\hat H_2$ and $Y$, we obtain the variational equation 
\begin{align}
    \label{SaddlePointEquation}
     \nu_n&=\frac{1}{\pi}\mathrm{Im}\frac{1}{v_n -i  \kappa_n  },\cr
     &\kappa_n \equiv\pi (\mathcal{P}\hat \nu)_n\equiv\pi \sum_m \mathcal{P}_{|n-m|}\nu_m,
\end{align}
where $\hat{\nu}$ denotes the matrix diagonal in the occupation basis, with elements $\nu_{n}$, and where we introduce 
the variational level hybridization, $\kappa_n$.  The
structure of this equation contains the key to its solution:
For $v_n=0$, the normalization $\sum_m \mathcal{P}_{|n-m|}=1$ implies that it is solved by $\kappa_n=1$. In the chosen units, this is ($\pi\times $) the density of states at the SYK band center. For finite
$v_n$, the summation over $m$ implements an effective average over the connected
states, which now carry random energy. In Appendix~\ref{app1} we show that the
average stabilizes the solution~\eqref{SaddlePointSolution},
\begin{align}
\label{SaddlePointSolution}
	&\kappa_n \simeq \kappa \Theta(C-|v_n|),\cr
    &(\kappa,C)
    =
    \begin{cases}
       (1,1),&\quad \delta<1\quad (\mathrm{I, II}),\cr
      \left(\delta^{-1},\delta\right),&\quad \delta>1\quad (\mathrm{III, IV}),
    \end{cases}
\end{align} 
where `$\simeq$' stands 
for equality up to corrections exponentially small in $\exp(-(v_n/\delta
)^2))$. We interpret this result as the spectral density of sites with energy
$v_n$ and decay rate $\kappa_n$ into neighboring sites. The latter is finite
for states below a threshold $|v_n|<C$. For $\delta>1$, the rate is given by the
 energy denominator
$\kappa\sim \delta^{-1}$ of neighboring sites. In the
opposite regime, $\delta<1$, the energy denominators of states $v_n\sim 1$ in
resonance with the SYK band width are of $\mathcal{O}(1)$, leading to the second line
in Eq.~\eqref{SaddlePointSolution}.

The saddle point solutions discussed thus far are distinguished for their diagonality
in all matrix indices. However, we now recall that the $z=0$ action is invariant
under Fock space uniform rotations Eq.~\eqref{ZereModeSymmetry}, implying that
uniformly rotated saddle point configurations $Y_n \to T Y_n T^{-1}$ are solutions,
too. (Technically, this follows from the cyclic invariance of the trace.) Next to
this uniform Goldstone mode, configurations $Y_n \to T_n Y_n T_n^{-1}$ with
site-diagonal rotations commutative with $\hat H_2$ are expected to cost the least
amount of action. With $Y_n=\pi\nu_n \sigma_3$, 
this makes $Y_n \to \pi\nu_n Q_n$, $Q_n=T_n \sigma_3 T_n^{-1}$ 
the effective degrees of freedom of the theory, and substitution into Eq.~\eqref{YAction} defines the Goldstone mode integral,
\begin{align}
 \label{QAction}
   &{\cal Z}=\int {\cal D}Q\, e^{- S[Q]},\\
   &S
=
-{\pi^2\over 2}
{\rm STr}( (\hat \nu\hat Q) \mathcal{P} (\hat\nu\hat Q) )
+
{\rm STr}\log\left(
z 
-\hat H_2 
+
i\pi\mathcal{P}(\hat \nu\hat Q)
\right),\nonumber
\end{align}
where $\hat{Q}$ again denotes the matrix diagonal in the occupation basis, with elements $Q_{n}$.
In the next two sections, we investigate what this integral has to say about
wave function statistics and Fock space localization, respectively.

\section{Spectral and wave function statistics} 
\label{sec:wave_function_statistics}

In this section, we explore the spectral and wave function statistics in
regimes I-III. The presumption is that wave functions are not yet localized and
correlated with each other. This should lead to Wigner-Dyson spectral
statistics and wave function moments reflecting the extended nature on the
subsets of Fock space corresponding to active or resonant sites. 

To test these hypothesis it is sufficient to consider the integral~\eqref{QAction} in
the presence of effectively infinitesimal explicit symmetry breaking $z$: besides the
sources, $j$, this parameter contains a frequency argument $\omega\sim D^{-1}$ of the
order of the exponentially small inverse many body level spacing in the case of
spectral statistics, Eq.~\eqref{gfK}, or the infinitesimal parameter $\eta$ in the
case of wave function statistics, Eq.~\eqref{gfIq}. On general grounds, we expect the
smallness in the `explicit' symmetry breaking in a Goldstone mode integral to lead to
singular contributions $\sim z^{-n}$ proportional in the inverse of the that
parameter after integration. (Inspection of the prefactors, $\eta^{q-1}$ in the definition of the wave function statistics shows that such singularities are actually required to obtain non-vanishing results.) These most singular contributions to the integral must come from the Goldstone mode fluctuations of least action, which are fluctuations homogeneous in Fock space,
\begin{align}
    \label{FockSpaceZeroMode}
    Q_n =T_n \sigma_3 T_n \to T\sigma_3 T^{-1}\equiv Q.
  \end{align}  
With
$[T,\mathcal P]=0$, the substitution $\hat \nu\hat Q\to \hat \nu Q$ into the action Eq.\eqref{QAction} leads to
\begin{align}
    \label{EffectiveActionWaveFunctions}
    S_0[Q,j]={\rm STr}\log\left(
z- j
-
\hat H_2 
+
i \hat \kappa Q
\right),
\end{align}
where we made the dependence $z\to z-j$ of the action on the sources $j\equiv j_{I,n}$ required to calculate moments via Eq.~\eqref{sourceI} explicit again, and we noted $\pi\mathcal{P}\hat \nu=\hat \kappa$.

Before proceeding, we note that the structure of this action is identical to that describing the Rosenzweig-Porter
model --- a single random matrix of dimension $D$ containing Gaussian distributed
disorder on the matrix diagonal~\cite{Ossipov}. 
An important difference is, however, that the diagonal disorder in the latter is uncorrelated, while 
the Fock-space diagonal disorder induced by $\hat H_2$ is highly correlated. 
As a consequence, 
the effective action for the Rosenzweig-Porter model only allows for homogenous saddle point solutions~\cite{Ossipov,NEE_SYK}, 
while here we encounter solutions that become inhomogeneous in Fock 
space once on-site disorder exceeds the $\hat H_4$ band width.
The inhomogeneity accounts for a site-dependent broadening $\kappa_n$, induced by correlations in the disorder amplitudes, 
and 
also manifests in a separation into regimes II/III of the regime of non ergodic extended states.
In the following, we discuss what this reduction of the model has to say about spectral and wave function statistics. 

\subsection{Spectral statistics} 
\label{sub:spectral_statistics}

To obtain a prediction for spectral correlations based on the representation Eq.~\eqref{QAction} with Fock space zero mode, we consider the
correlation function~\eqref{K}, represented through matrix integral Green
functions as in Eq.~\eqref{IqGtreenFunctions} and Eq.~\eqref{gfK}. To compute these
quantities from the effective theory, we need to expand the action
Eq.~\eqref{EffectiveActionWaveFunctions} to lowest order in the parameter
$\omega/\kappa\sim 1/D$, and to second order in the sources~\eqref{sourceK}. The straightforward $\omega$-expansion yields (cf. Eq.~\eqref{app_eq:FrequencyAction})
\begin{align}
  \label{ErgodicFrequencyAction}
    S_\omega[Q]\equiv -i \frac{\pi\nu (\omega+i\eta)}{2}\,\STr(Q\sigma_3),
\end{align}
where $\nu$ is the zero energy density of states, Eq.~\eqref{DoSAverage}. What
remains, is the source differentiation and the integration over the matrix $Q$. To get some intuition for the integral, notice that the non-linear degree of freedom $Q=T\sigma_3 T^{-1}$ affords a representation, $Q=U Q_0 U^{-1}$, where $U$ contains various compact angular variables (cf. Appendix~\ref{app4}), and 
\begin{align}
     \label{QStructure}
     Q_0=\left(\begin{matrix}
\cos\hat{\theta} & i\sin\hat{\theta} \\
-i\sin\hat{\theta}  & -\cos\hat{\theta} 
\end{matrix}\right),
   \end{align}   
a rotation matrix in causal space. Diagonal in super-space, this matrix is parameterized in terms of the two `Bogolubov' angles $\hat \theta=(i\theta_{\mathrm{b}},\theta_\mathrm{f})^T$, where $\theta_\mathrm{f}\in [0,\pi]$ is a compact rotation variable, and $\theta_\mathrm{b}\in \Bbb{R}^+$ a non-compact real variable. This representation reveals the geometry of the integration manifold as the product of a sphere ($\theta_\mathrm{f}$) and a hyperboloid ($\theta_\mathrm{f}$) (coupled by variables contained in $U$.) Where the physics of non-perturbative structures in spectral and wave function statistics, and localization is concerned, the most important player is the non-compact variable, $\theta_\mathrm{b}$, as only this one has the capacity to produce singular results. Heuristically, one may think of the model reduced to its dependence on this variable as a non-compact version of a Heisenberg-model, containing hyperboloidal, rather than compact spins as degrees of freedom. 

Referring for details of the source differentiation and the subsequent integration over the matrix $Q$~\cite{Efetov} to Appendix~\ref{app4}, the above reduction of the model yields the GUE spectral correlation
function~\eqref{GUESpectralStatistics} for the spectral statistics on scales of the
many body level spacing in regimes I-III. With increasing energies, the assumption of
homogeneity of fluctuations in Fock space breaks down (cf.\ the next section) beyond a
`Thouless energy' whose value depends on the specific observable under
consideration~\cite{fn6}. However, the detailed investigation of Thouless thresholds for the present model is beyond the scope of the paper.

\subsection{Wave function statistics} 
\label{sub:wave_function_statistics}

 In the same manner, we may consider the local moments of wave functions
Eq.~\eqref{Iq}, represented via Green functions Eq.~\eqref{IqGtreenFunctions}, and
obtained from the matrix integral through Eq.~\eqref{gfIq}. A key feature of this
expression is that it contains a limit $\lim_{\eta\to 0}\eta^{q-1}(\dots)$ the factor
$\eta^{q-1}$ must thus be compensated for by an equally strong singularity
$\eta^{1-q}$ from the integral, where $\eta$ couples through $z = i \eta \sigma_3$.
Setting $\omega=0$ in Eq.~\eqref{ErgodicFrequencyAction} and integrating over the
functional differentiated in sources (a calculation detailed in Appendix~\ref{app4}),
then yields the moments~\eqref{MomentsI}--\eqref{IqRegimeII-III}.

The support of wave functions in regimes II and III is different (as indicated by the different value of $D_\mathrm{res}$ in Eqs.~\eqref{IqRegimeII-III}), while the DoS Eq.~\eqref{DoSAverage} assumes the same value.
The reason for this is that, in regime II, there is no distinction between active and resonant sites: there are $\sim D/(\sqrt{N}\delta)$ active sites contributing with unit weight to the DoS. By contrast, in III, the dominant contribution to the DoS comes from the smaller number of $D_\mathrm{res}\sim D/(\sqrt{N}\delta^2)$ resonant sites, with sharply peaked spectral weight $\sim \delta$, $\nu\sim D_\mathrm{res}\delta\sim D/(\sqrt{N}\delta)$.

\subsection{Comparison to numerics} 
\label{sub:comparison_to_numerics}

To numerically check the predictions for the statistics of many body wave
functions and spectra,
we calculated eigenfunctions and spectrum
from exact diagonalization of the Hamiltonian
$\hat{H} = \hat{H}_4 + \hat{H}_2$, see~\eqref{syk} and~\eqref{syk2n}, for $\{v_i\}$ obtained by diagonalizing~\eqref{syk2chi} as a one-body problem,
for $2N=22, 26, 30$ Majorana fermions
and varying values of $\delta$.
We kept $1/7$ of the total spectrum
and verified both
a nearly constant density of states
and that results remain unchanged
when we restrict to a smaller energy window.
From the selected eigenfunctions in the center of the band,
we calculated statistics of the moments of the wave function according to
Eq.~\eqref{Iq}.
The eigenfunctions are normalized in each definite parity subspace.
For the spectrum we compared
the numerical statistical distribution
with both Wigner-Dyson
and Poisson distributions by calculating the Kullback-Leibler divergence,
$KL\equiv D(P||Q) = \sum_k p_k \log(\frac{p_k}{q_k})$,
where $p_k$ is the spectral statistics from numerical data
and $q_k$ the respective distribution. 
In order to avoid level unfolding, we followed Ref.~\onlinecite{Huse07} and
studied the statistics of ratios of energy spacings, $r_j =
\mathrm{min}(\frac{s_j}{s_{j-1}}, \frac{s_{j-1}}{s_{j}})$, where $s_j = \epsilon_{j+1}
- \epsilon_{j}$ is the nearest neighbor spacing of the eigenenergies $\{\epsilon_{j}\}$.
The $q_k$ are then given by numerically integrating either
the Wigner-Dyson
or the Poisson distribution
for the variable {$r$ over each bin centered at $r_k$,
given by Ref.~\onlinecite{Bogomolny13}, 
\begin{align}
	P(r) = \begin{cases}
		  \frac{81\sqrt{3}}{2\pi} \frac{(r + r^2)^2}{(1 + r + r^2)^4}
		+ \delta P(r),
		\ \text{Wigner-Dyson (GUE)}, \\
		 \frac{2}{(1 + r)^2}, \ \text{Poisson,}
	\end{cases}
\end{align}
where $\delta P$ is a numerical correction given by 
$\delta P = \frac{2C}{(1 + r)^2}
	\left[(r + 1/r)^{-2} - c_2(r + 1/r)^{-3}\right]$, 
with $c_2 = 4(4-\pi)/(3\pi-8)$ and $C = 0.578846$ is obtained from fitting numerical
results in the GUE~\cite{Bogomolny13}.

In all figures the numerical values result from averaging over
eigenvectors and spectrum, taken from the band center and from both
parity sectors, of at least $1000$ independent realizations of the model. 
In computing the Kullback-Leibler divergence, the numerical distribution for $r_j$ is
obtained by splitting the interval $[0, 1]$ into $50$ bins of equal widths.


\section{Extended-to-localized transition}
\label{Ergodic_loc}

In regimes, II, III, the dominant contribution to the matrix integral at the lowest energies comes from homogeneous contributions $Q$. Upon approaching the localization threshold III/IV, inhomogeneous fluctuations $Q\to \hat Q=\{Q_n\}$ gain in importance and eventually destabilize the mean field theory. To describe this physics, we need an effective action generalized for inhomogeneous fluctuations, and more manageable than Eq.~\eqref{QAction}. 
We derive it in Appendix~\ref{app2} under the assumption that the sum over a large number of fluctuating terms represented by the term $\mathcal{P}(\hat \nu \hat Q)$ is largely self-averaging.  
 An expansion to lowest order in fluctuations around the homogeneous average then leads to the effective hopping action
\begin{align}
\label{QActionHopping}
S[Q]&=S_{\cal P}[Q]+S_{\omega}[Q],\cr
 &S_{\cal P}[Q] 
  =
  \frac{\pi^2}{2}\sum_{n,m} \nu_n \nu_{m} {\cal P}_{n,m}
     \mathrm{Str}(Q_nQ_{m}),
  \\
\label{QActionFrequency}
  &S_{\omega}[Q] 
  =
    - i\pi \sum_{n}  \nu_{n} \, 
\mathrm{Str} \left(z Q_n\right), 
\end{align} 
with $Q_n=T^{-1}_n\sigma_3T_n$ and `${\rm Str}$' traces only over internal degrees of freedom. 
Eqs.~\eqref{QActionHopping} and~\eqref{QActionFrequency} are the main result of this section. Depending on the value of $\kappa_n$ Eq.~\eqref{SaddlePointSolution}, this action describes the entire range from vanishing to large deformations $\hat H_2$. 
We next discuss what this action has to say on the ergodic-to-localization transition. 

The key player in this problem is the hopping term~\eqref{QActionHopping} where $Q$ matrices at $\hat H_4$-neighboring sites are coupled, subject to a weight which contains the local spectral densities. In analytic approaches to localization on high dimensional lattices, it is common to set these weights to unity. However, in view of the massive site-to-site fluctuations of $\nu_n$ we prefer not to make this assumption and work with a given realization $\{\nu_n\}$ for as long as possible. Approaching the transition from the localized side where the integration over $Q$'s is subject to only small damping $\nu_n$, the essential degrees of freedom are once again the non-compact variables, $\theta_\mathrm{b}$ contained in $Q_0$, Eq.~\eqref{QStructure}.

To better understand the significance of this structure, we write $Q_n Q_m=(Q_n  - \sigma_3)(Q_m- \sigma_3)+\sigma_3 Q_n + \sigma_3 Q_m -\mathds{1}$ to represent the hopping part of the action as
\begin{align*}
  S_{\cal P}[Q] 
  &=\pi\sum_n  \Gamma_n \Str(Q_n \sigma_3)\\
  &+  
  \frac{\pi^2}{2}\sum_{n,m} \nu_n \nu_{m} {\cal P}_{n,m}
     \mathrm{Str}((Q_n-\sigma_3)(Q_{m}-\sigma_3)),
\end{align*}
where $\Gamma_n\equiv \nu_n\sum_m {\cal P}_{n,m} \nu_m$. Consider a situation where the accumulate hopping weights $\Gamma_n$ out of site $n$ are small. In this case, large fluctuations of the non-compact angles, $\lambda_{{\rm b},n} \equiv \cosh(\theta_{\mathrm{b},n})$ dominate the functional integral. To understand the consequences, we note that the measure of the $Q$-integration in the angular representation is given by~\cite{Efetov}
\begin{align*}
\int dQ=\int dU \int_{-1}^1 d\lambda_{\rm f} \int_1^\infty d\lambda_{\rm b}\frac{1}{(\lambda_{\rm b}- \lambda_{\rm f})^2},
\end{align*}
where $\lambda_{\rm f}=\cos(\theta_\mathrm{f})$. For small typical values $\Gamma\sim
\Gamma_n\ll 1$, the exponential weighs effectively cut off the integration over
$\lambda_{\rm b}$ at $\sim \Gamma^{-1}\gg 1$. Individual terms in the second line of the
above representation of $S_\mathcal{P}$ are smaller than the accumulated weights in
the first line, and so the integral can be approached by perturbative expansion in
the hopping terms. As an example, consider the sixth order expansion indicated via the 
highlighted links in Fig.~\ref{fig:NoLoop}. Retaining only the information on the non-compact integrations 
$\lambda\equiv\lambda_{\rm b}$, the contribution with a loop (left) and that with doubly occurring links evaluate to
\begin{align}
\text{ loop: 
}&\int_1^{\Gamma^{-1}} \frac{d \lambda_1}{\lambda_1^2} \frac{d \lambda_2}{\lambda_2^2}\frac{d \lambda_3}{\lambda_3^2}\frac{d \lambda_4}{\lambda_4^2} \lambda_1^3 \lambda_2^3 \lambda_3^2\lambda_4^2\sim  \Gamma^{-6},\cr
\text{ no loop:
}&\int_1^{\Gamma^{-1}} \frac{d \lambda_1}{\lambda_1^2} \frac{d \lambda_2}{\lambda_2^2}\frac{d \lambda_3}{\lambda_3^2} \lambda_1^3 \lambda_2^5 \lambda_3^2\sim  \Gamma^{-7},
\end{align}
where the indices refer to the participating $Q$-matrices, $Q_{1\ldots 4}$.
This estimate shows that the contribution of loops in the perturbation expansion is suppressed. At the same time, the largeness of the individual contributions signals that infinite order summations are required. The effective medium approximation achieves this summation, loops excluded. The approximation is called `effective medium' because from the perspective of individual sites in Fock space the contribution of all hopping processes terminating at that site adds up to the influence of an effectively homogeneous background medium, transmissive or not depending on the strength of the couplings.

\begin{figure}
\centering
\includegraphics[width=5.5cm]{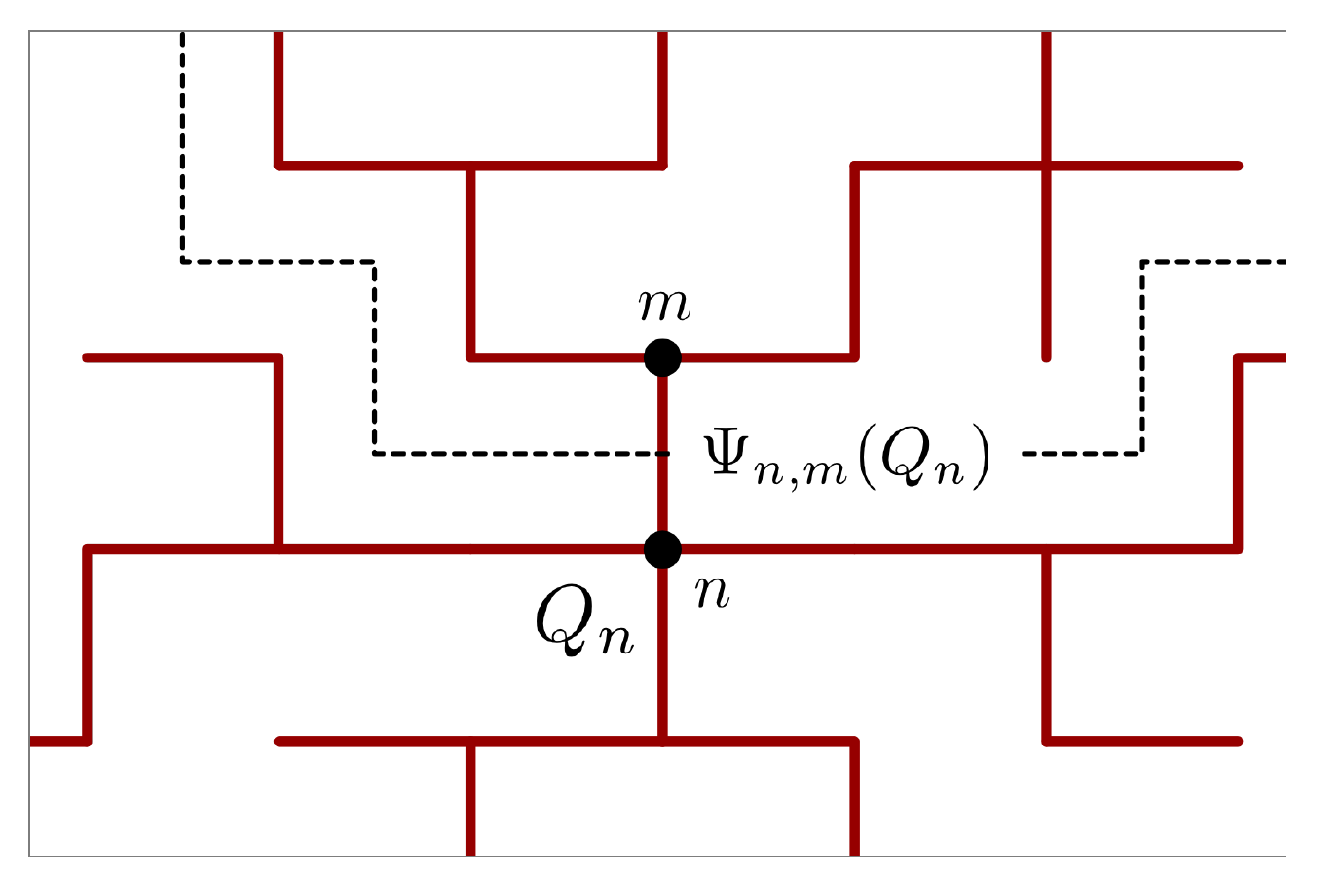}
\caption{\label{fig:EffectiveMedium} 
Idea of the effective medium approximation. Sites $n$ are connected to the stems of coral like structures, each labeled by a connected neighbor $m$, which represent the summation over all hopping terms excluding loops. The recursive nature of the structure allows for a self consistent resummation.
 }
\end{figure}

To see how this comes about, consider a site $n$ with local configuration $Q_n$ and
let $ 
\Psi_{n,m}(Q_n)=\int_{\text{coral}\,m,Q}DQ\, e^{-S[Q]},
$ be the contribution to the functional integrated over all
links connected to $n$ via the neighbor $m$, through the loopless coral like
structure indicated in Fig.~\ref{fig:EffectiveMedium}. The essence of the approximation is the recursion relation,
 \begin{align*}
 \Psi_{nm}(Q)&=\int dQ' N_{w_{nm}}(Q,Q')e^{-S_0(Q')}\prod_o  \Psi_{mo}(Q'),\cr 
 &N_{w}(Q,Q')=e^{w\,
     \mathrm{Str}(QQ')},
 \end{align*}
 where the product extends over all sites, $o$, connected to $m$ by hopping, $S_0(Q)\equiv S_{\omega\to i\delta}(Q)$ acts as a convergence generating factor, and we defined
\begin{align}
\label{eq:wnmDef}
w_{nm}\equiv\frac{\pi^2}{2} \nu_n \nu_{m} {\cal P}_{n,m},
\end{align}
for the coupling constants weighting the hopping kernel.
 If we now take the product $\Psi_n(Q)\equiv \prod_{m} \Psi_{n,m}(Q)$ (assuming self averaging in the sense that the fully integrated amplitude, $\Psi_n$ depends on the terminal site, $n$, but not on the detailed values of the $\mathcal{O}(N^4)$ neighbor amplitudes), the equation assumes the form, 
\begin{align*}
 \Psi_{n}(Q)&=\prod_{m}\int dQ' N_{w_{n,m}}(Q,Q')\Psi_m(Q'),
 \end{align*}
 where the presence of the convergence generator $\exp(-S_0)$ is left implicit.
 In the deeply localized regime, $N_{w_{n,m}}\approx 1$, the integral decouples, and $\Psi_n=1$ is a solution by supersymmetry (i.e.\ the unit normalization of all source-less integrals in the present formalism). This suggests~\cite{Efetov} a linearization, $\Psi_n(Q)=1-\Phi_n(Q)$, where the emergence of a non-trivial solution $\Phi_n$ is taken as a criterion for the localization transition. Substituting this ansatz into the equation, and again using supersymmetry, $\prod_{m}\int dQ' N_{w}(Q,Q')=1$, we obtain
\begin{align}
\label{eq:EffectiveMediumLinearized}
 &\Phi_{n}(Q)=\sum_{m}\int dQ' N_{w_{nm}}(Q,Q')\Phi_m(Q').
 \end{align}
This is a linear integral equation governed by a random lattice structure in Fock
space via the couplings $w_{nm}$ and an internal structure encoding the randomness
of the $\hat H_4$ system via the $Q'$-integrals. Although the integral equation may look
helplessly complicated, progress is possible 
recalling our previous observation: 
we again have a situation where the $Q$ integrations extend over wide parameter
intervals such that the leading non-compact variable is the key player. Assuming that
the solutions depend on the non-compact variable as $\Phi(Q)\to \Phi(t), t\equiv
\log(\lambda_1/\delta)$, and referring to Ref.~\cite{Zirn86} for details of the
integration over remaining variables, the reduction of
Eq.~\eqref{eq:EffectiveMediumLinearized} to the regime of interest, $t\ll 0$,
$w_{mn}\ll 1$ reads
\begin{align}
  \label{eq:LKernelDef}
  \Phi_n(t)&=\sum_m \int dt' L_{w_{mn}}(t-t') \Phi_m(t'),\\ \nonumber
  & L_w(t)=\left(\frac{w}{2\pi}\right)^{1/2}e^{-w \cosh(t)+\frac{t}{2}}\left(w \cosh t+\frac{1}{2}\right).
\end{align}
Ref.~\cite{Zirn86} contains a pedagogical discussion of the solution of the homogeneous variant $w_{mn}=\mathrm{const}.$ of this
equation, including the somewhat subtle issue of boundary conditions. It turns out that the key to the stability of the localized solution,
$\Psi=1$ lies in the spectrum of the linear kernel $\{L_{w_{mn}}(t-t')\}$: a spectrum with lower bound $\epsilon>1$ means that perturbations $\delta \psi$ will grow under the application of the linearized kernel, signifying destabilization of the null solution $\Psi=1$. We thus declare the existence of a minimal eigenvalue $\epsilon=1$ as a delocalization criterion. 
Due to translational invariance in $t-t'$ eigenstates are of the form
$e^{\theta(t-t')}\Phi_n$, where the coefficients are determined by the reduced equation, $\Phi_n= \sum_m L_{\theta,nm}\Phi_m$, 
with 
\begin{align*}
  L_{\theta,nm}&= \int_{-\infty}^\infty dt L_{w_{mn}}(t)e^{-\theta t}. 
\end{align*}
Substitution of the kernel in Eq.~\eqref{eq:LKernelDef} followed by 
differentiation in $\theta$ shows that the positive matrix $L_{\theta,nm}$ assumes its smallest values at $\theta=1/2$, 
and the straightforward integration at that value defines the matrix, 
\begin{align*}
  L_{nm}&\equiv L_{\frac{1}{2},nm}\cr
  &=\left(\frac{w_{nm}}{2\pi}\right)^{1/2} \int dt\,e^{-w_{nm} \cosh t}\left(w_{nm} \cosh t+\frac{1}{2}\right)\cr
&\simeq  \left(\frac{w_{nm}}{2\pi}\right)^{1/2}\log\left(\frac{2}{w_{mn}}\right).
\end{align*}
We thus arrive at the eigen equation 
\begin{align}
\label{eq_dc}
  \Phi_n
&=
{2\sqrt{\pi}\over \sqrt{\rho}}
\sum_{|n-m|=4} a_{nm}
\Phi_m,\cr 
&\qquad 
a_{nm}
=
\sqrt{\nu_n\nu_m}
\log\left(
{\rho\over (2\pi)^2\nu_n\nu_m}
\right), 
\end{align}
where the sum extends over $Z\equiv\binom{N}{4}$ sites in Hamming distance four to
 the reference site $n$~\cite{hamming}, and we recall that $\rho\equiv
 \binom{2N}{4}$. We read Eq.~\eqref{eq_dc} as an equation for the existence of a unit
 eigenvalue whose solvability depends on the value of $\delta$ determining the local
 density of states, $\nu_n$. In Appendix~\ref{app_loccrit} we show that the summation
 in this equation is dominated by resonant sites, and how this simplifies its
 logarithmic dependence.  Once again using the self averaging feature to replace the
 sum by an average over the distribution of $\nu_m$, we find that Eq.~\eqref{eq_dc}
 has a solution for $\delta=\delta_c$ determined by  the criterion
 Eq.~\eqref{app_dcLW}. In the limit $N\gg 1$ the latter simplifies to
 Eq.~\eqref{deta_c_d}.  However, as discussed in
 section~\ref{sub:strong_localization}, the numerical data for small values
 $N=\mathrm{10^1}$ responds sensitively to such approximations and improved agreement
 is obtained by working with the solution Eq.~\eqref{app_dcLW}.

\section{Discussion}
\label{discussion}

In this paper, we have presented a first principles analysis of Fock space
localization in the Majorana SYK$_{4+2}$ model, describing a competition of two-body
interaction and one-body potential. Within this setting, we provided a complete
description from an ergodic regime, over an intermediary regime of non-ergodic extended
states to the localized phase, all formulated in the eigenbasis of the one-body
Hamiltonian. 
Our main results are the identification of the MBL transition point, and the quantitative characterization of wave functions, particularly in regimes where they are neither localized nor trivially extended.

We compared the analytical results to numerical data for systems of size $N=11-15$ without fitting parameters. For systems of this size, the intermediate regime II is too narrow for a reliable comparison. However, in the ergodic regimes I and the strongly non-ergodic regime III we obtained  good agreement between analytical and numerical results. The finite size scaling of both wave function
and spectral statistics revealed an Anderson transition at a critical point which agreed with the
theoretically predicted value up to a size-independent numerical constant of
${\cal O}(1)$. In view of the numerous large $N$ approximations involved in the construction of the theory, we consider this a reassuring test for the applicability of localization theory on high dimensional lattices to realistic systems. 

Conceptually, the main contribution of the present work is an analytical description  which actually is not more complex than theories for phenomenological models of MBL. The high coordination number of the microscopic Fock space gave the system self averaging properties facilitating its analytic description. The resulting theory was tested for small sized systems $N=\mathcal{O}(10)$. However, it is expected to work the better the larger $N$, while the situation with computers is the other way around. On this basis, one may be cautiously optimistic that the concepts discussed here may become building blocks for the description of more complex MBL problems, including those with spatial structure.

\acknowledgements Discussions with A.~D.~Mirlin, K.~Tikhonov, D.~A.~Huse,
A.~Kamenev, and H.~Wang are gratefully acknowledged. 
F.~M and T.~M.~acknowledge financial support by Brazilian agencies CNPq and FAPERJ. 
The work of M. T. was partially supported by JSPS KAKENHI Grants JP17K17822, JP20K03787, and JP20H05270.
Work funded by the Deutsche Forschungsgemeinschaft (DFG, German Research Foundation) - 
Projektnummer 277101999 - TRR 183 (project A03).
Part of the numerical computation in this work was carried out
at the Supercomputer Center, ISSP, University of Tokyo.

\begin{appendix}

\section{Derivation of the action \eqref{YAction}} 
\label{sec:derivation_of_the_action}

We here derive the action Eq.~\eqref{YAction} from the averaged functional~\eqref{SPsiQuartic}. We start by rewriting the quartic term as $(\bar \psi \hat X_a \psi)^2=\STr((\psi\bar \psi \hat X_a)^2)$. To decouple this nonlinearity, we multiply the functional with the unit normalized
Gaussian integral $1=\int DA \,\exp(-\frac{1}{2}\sum_a\STr(A_a\hat
X_a)^2)$, where $DA\equiv \prod_a dA_a$, and $A_a=\{A_{nn'}^{ss',\sigma\sigma'}\}$
are $4D$-dimensional matrices. A shift $A_a \to A_a + w\psi\bar \psi$ then removes the
quartic term, and the subsequent integration over $\psi$ leads to
\begin{align*}
    {\cal Z}[j]=\int DA\,e^{-\frac{1}{2}\sum_a\STr(A_a\hat X_a)^2-\STr\log(\hat G^{-1}+w\sum_a A_a)},
\end{align*}
where $\hat G^{-1}= z - \hat H_2$, and we changed $A_a\mapsto \hat X_a A_a \hat X_a$. 
We now observe that the nonlinear part of the action couples only to the combination
$\sum_a A_a$. This motivates the definition, $A_a=\frac{i}{\rho}(Y +  Y_a)$, where the factor of
$i$ is included for later convenience, and $\sum_a Y_a=0$. Adding a Lagrange
multiplier $\frac{i}{\rho}\sum_a\STr(Y_a \Lambda)$ to
enforce the constraint, we are led to consider the functional $\mathcal{Z}[j]=\int
DYD\Lambda \exp(-S[Y,\Lambda])$, with action
\begin{align*}
    S[Y,\Lambda]&=-\frac{1}{2\rho^2}\sum_a\STr((Y+Y_a) P_a(Y+Y_a)) \cr
    &+\frac{i}{\rho}\sum_a \STr(\Lambda Y_a)+\STr\log(\hat G^{-1}+ iw Y),
\end{align*}
where $\rho=\binom{2N}{4}$ and we defined the operator $\hat P_a B=\hat X_a B \hat X_a$. Note that $\hat P_a$
is self-inverse, $\hat P_a^2 B =\hat X_a^2 B \hat X_a^2 =B$, and hermitian in the
sense that $\STr(C\hat P_a B)=\STr(\hat P_a C B)$. We now do the Gaussian integrals
over $Y_a$ to obtain,
\begin{align*}
    S[Y,\Lambda]= - \, \STr\big(\frac{\rho}{2}\Lambda \mathcal{P}\Lambda+i\Lambda Y\big)+\STr\log(\hat G^{-1}+i w Y),
\end{align*}
where $\mathcal{P}=\frac{1}{\rho }\sum_a \hat P_a$. The Gaussian integration over $\Lambda$ 
may now be performed and after rescaling $Y\to \rho^{1/2}Y$, and defining $\gamma = w \rho ^{1/2}=\tfrac{J}{2}(2N)^{1/2}$ we obtain the action $S[Y]
=
-{1\over 2}
{\rm STr}( Y \mathcal{P}^{-1} Y )
+
{\rm STr}\log\left(
z 
-\hat H_2 
+
i\gamma Y
\right)$. In a final step, we perform a linear transformation $\mathcal{P}^{-1}Y\to Y$, and recall that in our units 
$J^2=2/N$ and $\gamma= 1$, to arrive at Eq.~\eqref{YAction}.

\section{The operator $\mathcal{P}$} 
\label{app_operator_P}

In this Appendix, we discuss the action of the operator $\mathcal{P}$ states
$|n\rangle\langle n|$ diagonal in the occupation number basis. To this end, note that
for a state $|n\rangle=|n_1,\dots,n_i,\dots,n_N\rangle$, the action of the
Majorana operator $\hat{\chi}_{2i}=c_i + c_i^\dagger$ produces the state
$|n_i\rangle \equiv \hat{\chi}_{2i}|n\rangle=|n_1,\dots,\bar
n_i,\dots,n_N\rangle$, where $\bar n$ is 0 for $n=1$, and vice versa. Similarly,
$\hat{\chi}_{2i-1}|n\rangle =i(-)^{n_i}|n_i\rangle$. Except for $n_i$
all other occupation numbers remain unchanged, and no superpositions of states are
generated. The adjoint action thus generates $\hat{\chi}_{2i}|n\rangle \langle
n|\hat{\chi}_{2i}=\hat{\chi}_{2i-1}|n\rangle \langle
n|\hat{\chi}_{2i-1}=|n_i\rangle \langle n_i|$, which we interpret as nearest
neighbor hopping in Fock space. Notice that
$(\hat{\chi}_{2i}\hat{\chi}_{2i-1})|n\rangle \langle
n|(\hat{\chi}_{2i-1}\hat{\chi}_{2i})=|n\rangle \langle n|$ leaves the state unchanged.

With these structures in place, it is straightforward to describe the action of
$\mathcal{P}|n\rangle\langle n|=\frac{1}{\rho}\sum_{a}\hat X_a |n\rangle\langle n|\hat
X_a$. The summation contains contributions changing the particle number $|n|$ by
$0,2$ and $4$. With $\mathcal{P}_{n,m}=\langle m| (\mathcal{P}|n\rangle\langle
n|)|m\rangle$, the diagonal contribution, $\mathcal{P}_0$ is obtained from the
$\binom{N}{2}$ terms of the structure $\hat{\chi}_{2i}\hat{\chi}_{2i+1}
\hat{\chi}_{2\beta}\hat{\chi}_{2\beta+1}$. Similar counting for the contributions changing $|n|$ by two and four 
gives the matrix elements stated in the main text,
\begin{align}
\label{app_PMatrixElements}
\mathcal{P}_0
=\frac{N(N-1)}{2\rho}, \quad \mathcal{P}_2=\frac{4(N-2)}{\rho},\quad \mathcal{P}_4=\frac{16}{\rho},
\end{align}
and it is verified that 
\begin{align}
&\sum_m 
\mathcal{P}_{m,n}
\nonumber\\
&=
\binom{N}{0}{N(N-1)\over 2\rho} 
 +
 \binom{N}{2}{4(N-2)\over \rho}
 +
 \binom{N}{4}{16\over \rho}
\nonumber\\
& =
 1.
  \end{align}

\section{Saddle point equations}
\label{app1}

In this Appendix we address the solution of the saddle point equation
Eq.~\eqref{SaddlePointEquation}. The non-trivial element in this equation is the quantity $\kappa_n\equiv \pi(\mathcal{P}\hat \nu)_n$ in the denominator. 
In terms of this quantity, Eq.~\eqref{SaddlePointEquation} becomes the simple algebraic equation~\eqref{SaddlePointSolution}. A closed yet site non-local equation for $\kappa$ is obtained by acting on Eq.~\eqref{SaddlePointEquation} with the operator $\mathcal{P}$,
\begin{align*}
     \kappa_n &= \sum_m \mathcal{P}_{|n-m|} \mathrm{Im}\frac{1}{v_m - i \kappa_m} \cr
     &=\sum_m \mathcal{P}_{|n-m|}\mathrm{Re}\int_0^\infty dt\, e^{iv_m t - \kappa_mt},
 \end{align*} 
where in the second line switch to a temporal Fourier representation to facilitate the treatment of the argument $v_m$. The solution of this equation relies on two conceptual elements, first the
 ansatz Eq.~\eqref{SaddlePointSolution} and second a replacement of the sum over the
 $\rho$ neighboring sites $m$ by a Gaussian average over energies $v_m$.
 Specifically, we note that up to corrections small in $N^{-1}$, the neighbor sites
 $m$ are separated by Hamming distance $4$ from $n$ and each change in $n_i$ changes 
 $v_n \mapsto v_n \pm 2v_i$. 
 This means that $v_m =v_n + v$,
 where we assume $v$ to be Gaussian distributed with width $\sqrt{4}2\delta=4\delta$.
 Substituting the ansatz $\kappa_m = \kappa \Theta(C-|v_m|)$ into the equation, and
 splitting the integral over $v$ into regions with $C-|v_m| = C-|v_n+v|$ smaller and
 larger than zero, respectively, we obtain after shifting $v\mapsto v-v_n$
\begin{align*}
    &\kappa_n 
    =\frac{1}{\sqrt{32\pi}\delta}\mathrm{Re}\int_0^\infty dt\, \,\\
    &\times \left(\int dv \,e^{-\frac{(v-v_n)^{2}}{32\delta^2}} +\int_{-C}^{C} dv\,e^{-\frac{(v-v_n)^2}{32\delta^2}}  \left(e^{-\kappa t}-1\right)\right)
    e^{i vt}.
\end{align*}
With $\mathrm{Re}\int_0^\infty dt\,e^{i vt}=\pi \delta(v)$, the first and the third term in the second line cancel out, and the $t$-integration of the second term gives
\begin{align}
     \kappa_n &=\frac{\sqrt{\pi}}{\sqrt{32}\delta}\int_{-C}^{C} dv\,e^{-\frac{(v-v_n)^2}{32\delta^2}} \frac{\kappa}{\pi(v^2+\kappa^2)},
 \end{align} 
where the notation emphasizes that the $\kappa$-dependent term effectively represents a $\delta$-function $\delta_\kappa(v) = \frac{\kappa}{\pi(v^2+\kappa^2)}$ in $v$, smeared over scales $\sim \kappa$. This expression defines the mean field amplitude $\kappa_n$ at site $n$ in dependence of the tolerance window $C$ for the energy $v_n$, and $\kappa$ itself. We now explore for which configurations $(C,\kappa)$ it represents a self consistent solution.

The details of this analysis depend on wether we work with weakly (I, II) or
strongly (III, IV) distributed on-site energies.

\noindent \emph{Strong on-site disorder $III, IV$}: Anticipating that all solutions satisfy $\kappa\ll 1$, the width of $\delta_\kappa(v)$ is much smaller than that of the Gaussian weight, $\delta$. The function $\delta_\kappa$ thus collapses the integral, and we obtain
\begin{align}
\label{app_kappa}
     \kappa_n = \frac{\sqrt{\pi}}{\sqrt{32}\delta}e^{-\frac{v_n^2}{32\delta^2}}.
 \end{align} 
 This is consistent with our ansatz with $C=2\delta$ and $\kappa\sim \delta^{-1}$.

\noindent \emph{Narrow on-site disorder $I, II$}: In this regime, we test for
the validity of the ansatz with $C=1$ and $\kappa=1$. First assume $|v_n|>1=C\gg
\delta$. In this case, the ansatz requires exponentially suppressed $\kappa$, the
$\delta_v$-function again becomes effective, and the integral collapses to $\kappa_n
= \frac{\sqrt{\pi}}{\sqrt{32}\delta}\exp(-\frac{v_n^2}{32\delta^2})$ consistent with
the assumed smallness of $\kappa$. Conversely, for $|v_n|<1=C$, the ansatz requires
$\kappa=1$. The function $\delta_\kappa=\delta_1$ is now much wider than the width of
the Gaussian, $\sim \delta$, and the integration boundaries can be extended to
infinity. Doing the integral, we obtain $\kappa_n\equiv \kappa=1/\kappa$, or $\kappa=1$, consistent
with Eq.~\eqref{SaddlePointSolution}.

\section{Effective matrix theory}
\label{app2}

In this appendix we discuss the derivation of Eqs.~\eqref{QActionHopping} and~\eqref{QActionFrequency} from Eq.~\eqref{YAction}. In Eq.~\eqref{YAction}, we substitute $Y\to \pi\hat \nu \hat Q$ with $Q_n=T_n \sigma_3 T_n^{-1}$. The expansion of the action in fluctuations then comprises three parts: the Gaussian weight, the expansion of the `${\rm Str} \log$' in site-to-site fluctuations, and the expansion of the `${\rm Str} \log$'
in small frequency arguments, $z$ (reflecting the non-commutativity, $[z,T_n]\not=0$.)

\noindent \emph{Gaussian weight:} A straightforward substitution yields
\begin{align}
     \label{app_FlucutationsGaussianWeight}
&-{1\over 2}
{\rm STr}(  Y \mathcal{P}  Y )\to -{\pi^2\over 2}
{\rm STr}( \hat \nu\hat Q \mathcal{P} (\hat \nu \hat Q) ) \cr
&=-\frac{\pi^2}{2}\sum_{nm}\nu_n \nu_m P_{|n-m|}\Str{Q_nQ_m},
 \end{align}
 where `${\rm STr}$' includes the Fockspace trace, while `${\rm Str}$' is only over internal degrees of freedom.

\noindent \emph{Fluctuation action:}
Substituting the ansatz into the `${\rm Str} \log$' and temporarily neglecting the frequency arguments, $z$, we obtain
\begin{align}
    &\STr\log(-\hat H_2+ i  \pi\mathcal{P}(\hat \nu\hat Q)) \cr
    &=\STr\log(-\hat H_2+ i  \hat T^{-1}\pi\mathcal{P}(\hat \nu\hat Q)\hat T) \cr
    &=\STr\log(-\hat H_2+i   \pi\mathcal{P}(\hat \nu\sigma_3 )+ i\pi  [\hat T^{-1}\mathcal{P}(\hat \nu\hat Q)\hat T-\mathcal{P}(\hat \nu\sigma_3)]) \cr
    &\simeq\STr \log(1+\pi^2\hat \nu \sigma_3 [\hat T^{-1}\mathcal{P}(\hat \nu\hat Q)\hat T-\mathcal{P}(\hat \nu\sigma_3)]) \cr
    &\simeq \pi^2\STr(\hat \nu \sigma_3 [\hat T^{-1}\mathcal{P}(\hat \nu\hat Q)\hat T-\mathcal{P}(\hat \nu\sigma_3)]) \cr
    &=\pi^2 \STr(\hat \nu \hat Q\mathcal{P}(\hat \nu\hat Q)),
\end{align}
identical to $(-2\times)$ the Gaussian weight. 
In the second line we used the cyclic invariance $\STr \log(\dots)=\STr\log(\hat T^{-1} (\dots)\hat T)$, and in the fourth the saddle point equation $(-\hat H_2 +i \pi \mathcal{P}(\hat \nu\sigma_3))^{-1}=-i\pi\hat  \nu \sigma_3$.

\noindent \emph{Frequency action:} In a similar manner, we obtain
\begin{align}
  \label{app_eq:FrequencyAction}
    &\STr\log(-\hat H_2+ i  \pi\mathcal{P}(\hat \nu\hat Q)+z) \cr
    &\simeq \STr\log(\hat T(-\hat H_2+ i  \pi\mathcal{P}(\hat \nu\sigma_3))\hat T^{-1}+z) \cr
    &=\STr\log(-\hat H_2+ i  \pi\mathcal{P}(\hat \nu\sigma_3)+\hat T^{-1}z\hat T) \cr
    &\simeq -i\pi\STr(\hat \nu \sigma_3 \hat T^{-1}z \hat T)=
    -i\pi\STr(\hat \nu \hat Q z),
\end{align}
where in the second line, we neglected local fluctuations $ P(\hat \nu\hat T\sigma_3 \hat T^{-1})\simeq \hat T P(\hat \nu\sigma_3)\hat T^{-1}$, in the third used cyclic invariance, and in the fourth the saddle point condition. 

Combining terms, we obtain the effective action~\eqref{QActionHopping}.


\section{Wave function and spectral statistics from matrix model}
\label{app4}

In this section we provide details on the computation of wave-function and spectral statistics in the deformed $\hat H_4$ model. The starting point for both statistics is Eq.~\eqref{EffectiveActionWaveFunctions}, with sources $j=J_K$ or $J=J_{I,n}$, respectively, given in Eq.~\eqref{sourceK}.
Using the commutativity $[T,\hat H_2]=0$ we represent the action as
\begin{align*}
     S[T]
=
{\rm STr\log}\left(
1
+
\hat G{\cal O}_T
\right)=\sum_{k=1}^\infty \frac{(-1)^k}{k}\STr(\hat G\mathcal{O}_T)^k,
 \end{align*} 
where 
${\cal O}_T\equiv T^{-1} \left[ z
- j(\alpha,\beta)\right] T$ 
is an operator in which we need to expand to the order required by the correlation function, and we have made the source contribution, $j(\alpha,\beta)$, to the matrix $z=\frac{\omega+i\eta}{2}\sigma_3$ explicit again. Concerning the resolvent, $\hat G^{-1} \equiv  i \hat \kappa\, \sigma_3 - \hat H_2$, we notice that fluctuation variables commute through the real part of $\hat G$, and keep only $i\,\mathrm{Im}\hat G=-i\pi\hat \nu$, with local components $\nu_n$ defined in Eq.~\eqref{SaddlePointEquation}.
Specifically, to zeroth order in the sources, and first order in an expansion in $z \nu_n \sim \omega/\Delta$, the action assumes the form~\eqref{ErgodicFrequencyAction}.

For the computation of the spectral and wave function statistics, we need the expansion in sources to first order in $\beta$ and higher orders in $\alpha$. With the above definitions, the expansion of the action assumes the form
\begin{align}
S[T]= -\pi
\sum_{k=1}^\infty
\left(-i\nu_n\alpha\right)^k
\left(
{1\over k}
[ Q^{++}_{\rm bb} ]^k
+
\frac{\beta}{\alpha}
[Q^{++}_{\rm bb}] ^{k-1}Q^{--}_{\rm ff}
\right),
\end{align}
where in the terms $k>2$ we used the approximation 
$Q^{+-}_{\rm bf}Q^{-+}_{\rm fb}\simeq Q^{++}_{\rm bb}Q^{--}_{\rm ff}$ valid in the limit $\eta\to 0$ implied in the calculation of wave function moments~\cite{Efetov}. 
Doing the derivatives in the source parameters, we arrive at
\begin{align}
\label{app_sm_ev}
\partial_\alpha^{q-1} \partial_\beta{\cal Z}|_{\alpha,\beta=0}
&=
\left(-i\pi\nu_n\right)^q
q! \,
\langle 
\left[Q^{++}_{\rm bb} \right]^{q-1}
Q^{--}_{\rm ff}
\rangle,
\end{align}
where $\langle...\rangle=\int dQ \,e^{-S_z[Q]}(\dots)$. 

The remaining integral over the four-dimensional matrix $Q$ is conceptually straightforward but technically the hardest part of the calculation. Referring for details to Ref.~\onlinecite{Efetov}, we here review the main steps. 
 The starting point is a `polar coordinate' representation $Q=UQ_0U^{-1}$ with $Q_0$ defined in Eq.~\eqref{QStructure},
$\hat{\theta} ={\rm diag}(i\hat{\theta}_{\rm b}, \hat{\theta}_{\rm f})$ containing
compact and non-compact angles $0<\theta_{\rm f} <\pi$ and $\theta_{\rm b}>0$,
respectively~\cite{Efetov}. The matrix $U$ is block-diagonal in causal space and
contains four Grassmann variables $\eta^\pm, \bar\eta^\pm$, and two more commuting
variables $0\leq\phi,\hat{\chi}<2\pi$.
More specifically, $U={\rm diag}(u_1u_2,v)_{\rm ra}$, where
$u_2
={\rm diag}(
e^{i\phi},e^{i\hat{\chi} }
)_{\rm bf}$ 
and supermatrices $u_1=e^{-2\hat \eta^+}$, $v=e^{-2i\hat \eta^-}$,
generated by
$\hat{\eta}^\pm=
\left(\begin{smallmatrix}
0 & \bar{\eta}^\pm \\
-\eta^\pm & 0
\end{smallmatrix}\right)_{\rm bf}$. 
 In this representation, the matrix elements
entering the correlation function are given by $Q^{++}_{\rm bb}=\cosh\theta_{\rm
bb}(1-4\bar{\eta}^+\eta^+)$ and $Q^{--}_{\rm ff}=\cos\theta_{\rm
ff}(1-4\bar{\eta}^-\eta^-)$, and the integration measure reads $dQ={1\over
2^6\pi^2}{\sinh\theta_{\rm b}\sin\theta_{\rm f}\over (\cosh\theta_{\rm
b}-\cos\theta_{\rm f})^2} d\phi d\hat{\chi} d\theta_{\rm b}d\theta_{\rm f}
d\bar{\eta}^+d\eta^+d\bar{\eta}^-d\eta^-$~\cite{Efetov}. The essential advantage of
the polar representation is that the action only depends on the `radial variables'
$S_\eta[Q]=-i2\pi\nu(\omega+i\eta)(\cosh\theta_{\rm b}-\cos\theta_{\rm f})$. 

\noindent \emph{Wave function statistics:} In the calculation of the wave function moments, we may set $\omega=0$. The integration over the non-compact angle is then cut by the parameter $\eta$
at values 
$1\leq  
\cosh\theta_{\rm b}\lesssim 1/\eta$, while the
integration over the compact angles $\theta_\mathrm{f}$ is free. With this simplification, the integration over all variables except the non-compact one, $\theta$, becomes elementary, and one obtains~\cite{Efetov}
\begin{align}
&
G_{nn}^{+(q-1)} G^-_{nn}
=
2q(q-1) \left(-i\pi\nu_n\right)^q
\nonumber\\
&\times
\int_0^\infty 
d\theta_\mathrm{b} \,
\sinh\theta_\mathrm{b} \left( 
\cosh\theta_\mathrm{b} \right)^{q-2} 
 e^{-2\pi\nu\eta \cosh\theta_\mathrm{b}}. 
\end{align}
The final integral gives $(2\pi\nu \eta)^{1-q}q!$ and collecting all factors
we arrive at 
\begin{align}
     I_q=\frac{q!}{\nu^q}\sum_n
     \nu_n^q.
\end{align} 
 This result expresses the $q$th moment of the local wave function amplitudes through that of the local density of states individually averaged over $\hat H_4$ fluctuations. The energies $v_n$ at each individual site are obtained as sums of $N$ random coefficients $v_i$ (cf. Eq.~\eqref{syk2n}). For large $N$, this makes the sum self averaging, and we replace $I_q \to \langle I_q\rangle_v$ by its average over single particle energies, $v_i$. Using Eq.~\eqref{SaddlePointEquation}, we thus obtain
\begin{align*}
    I_q = \frac{(-)^{q-1}q}{(\pi\nu)^q}\sum_n ( \kappa_n)^q \left\langle \partial^{q-1}_{( \kappa_n)^2}\frac{1}{v_n^2+( \kappa_n)^2}\right\rangle.
\end{align*}
The evaluation of this expression now depends on which on-site disorder regime we are in. In regime I, $\delta<N^{-1/2}$, or $|v_n|<1$, the mean field broadening assumes the uniform value $\kappa=1$. In this case, the dependence of $I_q$ on site energies, $v_n$, is weak. This implies $\nu \simeq \frac{1}{\pi}\sum_n 1=D/\pi$. Doing the $\kappa$ derivatives, we obtain
\begin{align}
    \label{app_IqRegimeI}
    I_q= q! D^{1-q} ,\qquad \text{regime I},
\end{align}
which is the RMT result for a matrix of dimension $D$.

For larger disorder, only a fraction of sites have finite decay width. Using Eq.~\eqref{SaddlePointSolution} and assuming self averaging to replace the $n$-sum to an average over a distribution of site energies of width $\delta N$, the DoS is evaluated as
\begin{align*}
    \nu      
     &\simeq\frac{1}{\pi} 
     \frac{D}{\sqrt{2\pi N\delta^2}}\int_{-C}^C dv\,e^{-\frac{v^2}{2N\delta^2}}\frac{\kappa}{v^2+\kappa^2}
     \nonumber\\
      &\simeq \frac{1}{\pi}
     \frac{D}{\sqrt{2\pi N\delta^2}}\int_{-C}^C dv\,\frac{\kappa}{v^2+\kappa^2}
     \nonumber\\
      &=\frac{1}{\pi}
     \frac{2D}{\sqrt{2\pi N\delta^2}}
     \arctan(C/\kappa), 
\end{align*}
where in the second line we used that the distribution of energies is much wider than the tolerance window $C$. Substituting the values specified in Eq.~\eqref{SaddlePointSolution}, this leads to 
\begin{align}
    \label{app_DoSRegimesII-IV}
    \nu = c \frac{D}{\sqrt{N}\delta},
\end{align}
where $c$ is of order unity and the suppression relative to $\nu=c D$ in regime I accounts for the improbability to find resonant sites. 

In the same manner, we obtain 
\begin{align*}
    I_q 
    &\simeq
    \frac{(-)^{q-1} q}{(\pi\nu)^q}\frac{D}{\sqrt{2\pi N\delta^2}}\kappa^q\partial^{q-1}_{\kappa^2}\int_{-C}^Cdv\,e^{-\frac{v^2}{2N\delta^2}}\frac{1}{v^2+\kappa^2}\nonumber\\
    &\simeq\frac{(-)^{q-1} q}{(\pi\nu)^q}\frac{D}{\sqrt{2\pi N\delta^2}}\kappa^q\partial^{q-1}_{\kappa^2}\int_{-C}^Cdv\,\frac{1}{v^2+\kappa^2}\nonumber\\
&=2\frac{(-)^{q-1} q}{(\pi\nu)^q}\frac{D}{\sqrt{2\pi N\delta^2}}\kappa^q\partial^{q-1}_{\kappa^2} {1\over \kappa} \arctan(C/\kappa)\nonumber\\
&\simeq 2\frac{(-)^{q-1} q}{(\pi\nu)^q}\frac{D}{\sqrt{2\pi N\delta^2}}\kappa^q\partial^{q-1}_{\kappa^2} {1\over \kappa}\nonumber\\
&=\frac{1}{(\pi\nu)^q}\frac{D}{\sqrt{2\pi N\delta^2}} \frac{2q(2q-3)!!}{(2\kappa)^{q-1}},
\end{align*}
where `$\simeq$' here means equality up to some constant $c\sim{\cal O}(1)$. 
Insertion of Eq.~\eqref{app_DoSRegimesII-IV} leads to Eq.~\eqref{MomentsIIAndIII}.
Using Eq.~\eqref{SaddlePointSolution}, we finally obtain
\begin{align}
    \label{app_IqRegimeII-III}
    q\gg 1:\qquad I_q= c^q q!\left( \frac{D}{\sqrt{N}} \right)^{1-q}\left\{
\begin{array}{ll}
    \delta^{q-1},& \text{II},\cr
    \delta^{2(q-1)},& \text{III}.
\end{array}
    \right.
\end{align}
Finally, for a quantitative comparison to numerical simulations in regime ${\rm III}$ without fitting parameter
we trace all constants $c\sim{\cal O}(1)$ in $\nu$ and $I_q$.
Noting that in regime ${\rm III}$ we can substitute $\arctan(C/\kappa)=\pi/2$
 we arrive at, 
\begin{align}
\label{app_I_q_sc}
I_q 
&= 
{q(2q-3)!!\over (2\pi \nu\kappa)^{q-1}}
=
{q(2q-3)!!\over \delta^{2(1-q)} } 
\left({\pi D\over  4\sqrt{N}}\right)^{1-q},\qquad \text{III}
\end{align}
where in the second equality we used Eq.~\eqref{app_kappa} for $\kappa$.

\noindent {\it Level-statistics:} For the level statistics we need to keep finite
$\omega$, and differentiate the functional to first order in $\alpha$ and $\beta$ (Eq.~\eqref{gfK}). 
Application of Eq.~\eqref{app_sm_ev} then leads to~\cite{Efetov}
\begin{align}
K(\omega)
=
{1\over 2}{\rm Re}
&\int_0^\infty d\theta_{\rm b}
\int_{-\pi/2}^{\pi/2} d\theta_{\rm f}\,
\nonumber\\
&\times
 \sinh\theta_{\rm b}
\sin\theta_{\rm f}
e^{i\pi\nu\omega(\cosh\theta_{\rm b}-\cos\theta_{\rm f})}, 
\end{align}
where $\theta_{\rm b}$ and $\theta_{\rm f}$ are the non-compact bosonic and compact fermionic angle, respectively. 
 These integrals can be carried out in closed form, and yield the two-point correlation function of the Gaussian Unitary Ensemble~\eqref{GUESpectralStatistics}.

\section{Localization criterion}
\label{app_loccrit}

In this Appendix we demonstrate how the solution of the eigenvalue Equation~\eqref{eq_dc} reduces to the criterion~\eqref{deta_c_d}. We write the sum as
\begin{align*}
  \Phi_n
&=
{2\sqrt{\pi}\over \sqrt{\rho}}
\sum_{|n-m|=4} a_{nm}
\Phi_m,\cr 
&\qquad 
a_{nm}
=
\sqrt{\nu_n\nu_m}
\log\left(
{\rho\over (2\pi)^2\nu_n\nu_m}
\right),
\end{align*} 
and make the self consistent assumption that the sum 
 over neighboring sites $m$ is dominated by resonant sites, and that the solution, $\Phi_n$, too, are peaked at those sites. Under these conditions it makes sense to consider a zeroth order approximation 
$a_{nm}\simeq a^0_{nm}\equiv \sqrt{\nu_n\nu_m}
 2\log\left(
\sqrt{\rho}/2\pi \nu_m
\right)
$,
neglecting  site-to-site fluctuations of the logarithm. In a final step we will refine the result by perturbation theory in 
 $\delta a_{nm}\equiv a_{nm}- a^0_{nm}=
\sqrt{\nu_n\nu_m}
\log\left(\nu_m/\nu_n\right)$. Making the replacement $a_{nm}\rightarrow a_{nm}^0$, we observe that the equation is solved by 
 $\Phi_n\propto \sqrt{\nu_n}$, provided that
\begin{align}
\label{EigenvalueConsistency}
1
&=
{4 \sqrt{\pi}\over \sqrt{\rho}}
\sum_m
\nu_m
\log\left(
\frac{\sqrt{\rho}}{2\pi \nu_m}
\right),
\end{align}
where the sum extends over the $Z\equiv \binom{N}{4}$ sites in Hamming distance $4$ to $n$ (i.e.\ the parameter $Z$ defines the effective coordination number of the Fock space lattice.)
We note that with the above eigenstates the first order perturbative correction to the unit eigenvalue Eq.~\eqref{EigenvalueConsistency} is given by $\langle \Phi|\delta \alpha |\Phi\rangle\propto\sum_{nm}\nu_n\nu_m  
\log(
\nu_n/\nu_m)=0$, which we take as a self consistent justification to work with the zeroth order approximation. Turning to the consistency equation for the eigenvalue, we again replace the sum over nearest neighbors by an average over their distribution of energies (cf.\ a similar operation in Appendix~\ref{app1}):
\begin{align*}
 &\sum_m \nu_m f(\nu_m) \simeq  Z\langle \nu(v)f(\nu(v))\rangle_v\simeq Z \frac{f\left(\frac{\sqrt{32}\delta}{\sqrt{\pi}}\right)}{\sqrt{32\pi}\delta} ,\cr
&\qquad \langle \dots \rangle_v =
\frac{1}{\sqrt{2\pi}4\delta}\int dv \,e^{-\frac{v^2}{32\delta ^2}}(\dots).
\end{align*}
Here,  the second equality is based on the observation that on the subset of active sites, $v<\delta$, where $\nu(v)$ is non-vanishing, and $\nu(v)=\frac{\pi }{\delta(v^2+\delta^{-2})}$ becomes a  $\delta$-function of width $\sim \delta^{-1}$ and height $\nu(0)=\pi/\kappa$ with $\kappa=\frac{\sqrt{\pi}}{\sqrt{32}\delta}$ (cf. Eq.~\eqref{app_kappa}). The integral   collapses to this resonance region, leading to the stated result. (Effectively, this is saying that only resonant sites contribute to the nearest neighbor sum.)

Application of this auxiliary identity to the eigenvalue equation Eq.~\eqref{EigenvalueConsistency} leads to 
\begin{align}
1
&=
{1\over \sqrt{2\rho}}
{Z\over \delta}
\log\left(
\sqrt{\frac{8\rho}{\pi}}\delta
\right),
\end{align}
 which
 is solved by
 \begin{align}
 \label{app_dcLW}
\delta_c
&=
{ Z\over\sqrt{2 \rho}}
W\left(
2Z\sqrt{\pi}  
\right),
\end{align}
with $W$ the Lambert-$W$ function.

For  $N\gg 1$, we may approximate $Z=\binom{N}{4}\simeq N^4/24$ and $\rho=\binom{2N}{4}\simeq (2N)^4/4!$. The asymptotic expansion for large   arguments, $W(x)\simeq \log (x)+\dots$ then leads to the estimate 
Eq.~\eqref{deta_c_d} in the main text.

\end{appendix}     


\begin{thebibliography}{99}

\bibitem{Anderson}
P.~W.~Anderson, \textit{Absence of Diffusion in Certain Random Lattices}, Phys. Rev. {\bf 109}, 1492 (1958).

\bibitem{BaskoAleinerAltshuler}
D.~Basko, I.~Aleiner, and B.~Altshuler,
\textit{Metal–insulator transition in a weakly interacting many-electron system
with localized single-particle states},
Ann. Phys. {\bf 321}, 1126 (2006).

\bibitem{Mirlin}
I.~V.~Gornyi, A.~D.~Mirlin, and D.~G.~Polyakov,
\textit{Interacting Electrons in Disordered Wires: Anderson Localization and
Low-$T$ Transport},
Phys. Rev. Lett. {\bf 95}, 206603 (2005).



\bibitem{spin1}
M.~Znidaric, T.~Prosen, and P.~Prelovsek, 
\textit{Many-body localization in the Heisenberg XXZ magnet in a random field},
Phys. Rev. B {\bf 77}, 064426 (2008).

\bibitem{Huse2010}
A.~Pal, D.~A.~Huse, 
\textit{The many-body localization phase transition},
Phys. Rev. B {\bf 82}, 174411 (2010).

\bibitem{Moore2012}
J.~H.~Bardarson, F.~Pollmann, and J.~E.~Moore, 
\textit{Unbounded Growth of Entanglement in Models of Many-Body Localization},
Phys. Rev. Lett. {\bf 109}, 017202 (2012).

\bibitem{spin5}
M. Serbyn, Z. Papic, and D. A. Abanin,
\textit{Universal Slow Growth of Entanglement in Interacting Strongly Disordered Systems},
Phys. Rev. Lett. {\bf 110}, 260601 (2013).

\bibitem{spin3}
J.~A.~Kj\"all, J.~H.~Bardarson, and F.~Pollmann, 
\textit{Many-Body Localization in a Disordered Quantum Ising Chain}, 
Phys. Rev. Lett. {\bf 113}, 107204 (2014).

\bibitem{spin2}
K.~Agarwal, S.~Gopalakrishnan, M.~Knap, M.~M\"uller, and E.~Demler, 
\textit{Anomalous Diffusion and Griffiths Effects Near the Many-Body Localization Transition},
Phys. Rev. Lett. {\bf 114}, 160401 (2015).

\bibitem{ImbriePRL2016}
J.~Z.~Imbrie,  
\textit{Diagonalization and Many-Body Localization for a Disordered Quantum Spin Chain},
Phys. Rev. Lett. {\bf 117}, 027201 (2016). 

\bibitem{ImbrieJStatPhys2016}
J.~Z.~Imbrie,  
\textit{On Many-Body Localization for Quantum Spin Chains}
J.~Stat.~Phys. {\bf 163}, 998 (2016).

\bibitem{TorresSantos17} 
E.~J.~Torres-Herrera, and L.~F.~Santos,
\textit{Extended nonergodic states in disordered many-body quantum systems},
Ann. Phys. {\bf 529}, 1600284 (2017).

\bibitem{Laflorencie19}
N.~Macé F.~Alet, N.~Laflorencie,
\textit{Multifractal Scalings across the Many-Body Localization Transition},
Phys. Rev. Lett. {\bf 123}, 180601 (2019).





\bibitem{AltshulerGefenKamenevLevitov97}
B.~L.~Altshuler, Y.~Gefen, A.~Kamenev, and L.~S.~Levitov,
\textit{Quasiparticle Lifetime in a Finite System: A Nonperturbative Approach},
Phys. Rev. Lett. {\bf 78}, 2803 (1997).

\bibitem{Silvestrov97}
 P.~G.~Silvestrov, \textit{Decay of a Quasiparticle in a Quantum Dot: The Role of Energy Resolution}, 
Phys. Rev. Lett. {\bf 79}, 3994 (1997).

\bibitem{Silvestrov98} 
P.~G.~Silvestrov, \textit{Chaos thresholds in finite Fermi systems}, Phys. Rev. E {\bf 58}, 5629 (1998).

\bibitem{GornyMirlinPolyakov16}
I.~V.~Gornyi, A.~D.~Mirlin, and D.~G.~Polyakov,
\textit{Many-body delocalization transition and relaxation in a quantum dot},
Phys. Rev. B {\bf 93}, 125419 (2016).

\bibitem{GornyMirlinPolyakovBurin17}
I.~V.~Gornyi, A.~D.~Mirlin, D.~G.~Polyakov, A.~L.~Burin,
\textit{Spectral diffusion and scaling of many-body delocalization transitions}, 
 Annalen der Physik (Berlin) {\bf 529}, 1600360 (2017).
 
 
\bibitem{RubioAbadal19}
 A.~Rubio-Abadal, J.-Y.~Choi, J.~Zeiher, S.~Hollerith, J.~Rui, I.~Bloch, C.~Gross,
 \textit{Many-body delocalization in the presence of a quantum bath},
 Phys. Rev. X {\bf 9}, 041014 (2019).

\bibitem{Choi16}
J.-Y.~Choi, S.~Hild, J.~Zeiher, P.~Schau\ss, A.~Rubio-Abadal, T.~Yefsah, V.~Khemani, D.~A.~Huse, I.~Bloch, C.~Gross
 \textit{Exploring the many-body localization transition in two dimensions},
Science {\bf 352}, 1547 (2016).

\bibitem{Schreiber15}
M.~Schreiber, S.~S.~Hodgman, P.~Bordia, H.~P.~L\"uschen, M.~H.~Fischer, R.~Vosk, E.~Altman, U.~Schneider, I.~Bloch, 
\textit{Observation of many-body localization of interacting fermions in a quasi-random optical lattice}, 
Science {\bf 349}, 842 (2015).

\bibitem{kai18}
K.~Xu, J.J.~Chen, Y.~Zeng, Y.R.~Zhang, C.~Song, W.~Liu, Q.~Guo, P.~Zhang, D.~Xu, H.~Deng, K.~Huang, H.~Wang, X.~Zhu, D.~Zheng, H.~Fan,
\textit{Emulating Many-Body Localization with a Superconducting Quantum Processor},
Phys. Rev. Lett. {\bf 120}, 050507 (2018).

\bibitem{roushan17}
P.~Roushan, C.~Neill, J.~Tangpanitanon, V.~M.~Bastidas, A.~Megrant, R.~Barends, Y.~Chen, Z.~Chen, B.~Chiaro, A.~Dunsworth, A.~Fowler, B.~Foxen, M.~Giustina, E.~Jeffrey, J.~Kelly, E.~Lucero, J.~Mutus, M.~Neeley, C.~Quintana, D.~Sank, A.~Vainsencher, J.~Wenner, T.~White, H.~Neven, D.~G.~Angelakis, J.~Martinis,
\textit{Spectroscopic signatures of localization with interacting photons in superconducting qubits},
Science {\bf 358}, 6367 (2017).


\bibitem{NonE_Ex} 
A.~De~Luca, B.~L.~Altshuler, V.~E.~Kravtsov, and A.~Scardicchio,
\textit{Anderson Localization on the Bethe Lattice: Nonergodicity of Extended
States},
Phys. Rev. Lett. {\bf 113}, 046806 (2014).

\bibitem{Biroli18} 
G.~Biroli, and M.~Tarzia,
\textit{Delocalization and ergodicity of the Anderson model on Bethe lattices},
arXiv:1810.07545.

\bibitem{Mirlin1} 
K.~S.~Tikhonov, and A.~D.~Mirlin,
\textit{Statistics of eigenstates near the localization transition on random
regular graphs},
Phys. Rev. B \textbf{99}, 024202 (2019). 

\bibitem{TikhonovMirlin19_2} 
K.~S.~Tikhonov, and A.~D.~Mirlin,
\textit{Critical behavior at the localization transition on random regular
graphs},
Phys. Rev. B {\bf 99}, 214202 (2019).

\bibitem{faoro} 
L.~Faoro, M.~Feigel'man, and L.~Ioffe,
\textit{Non-ergodic extended phase of the Quantum Random Energy model},
Ann. of Phys. \textbf{409}, 167916 (2019)

 \bibitem{RPmodel1} 
V.~E.~Kravtsov, I.~M.~Khaymovich, E.~Cuevas, and M.~Amini,
\textit{A random matrix model with localization and ergodic transitions},
New Journal of Physics {\bf 17}, 122002 (2015).

\bibitem{LeyronasSilvestrovBeenakker2000}
X. Leyronas, P.G. Silvestrov, C.W.J. Beenakker,
\textit{Scaling at the chaos threshold in an interacting quantum dot},
 Phys. Rev. Lett. {\bf 84}, 3414 (2000).

\bibitem{SYK1}
S.~Sachdev and J.~Ye,
\textit{Gapless spin-fluid ground state in a random quantum Heisenberg magnet},
Phys. Rev. Lett. {\bf 70}, 3339 (1993).

\bibitem{SYK2}
A.~Kitaev, http://online.kitp.ucsb.edu/online/ entangled15/kitaev/ .... /kitaev2/ (Talks at KITP on April 7th and May 27th 2015).

\bibitem{SYK_GG}
A.~M.~Garc\'ia-Garc\'ia, B.~Loureiro, A.~Romero-Berm\'udez, and M.~Tezuka, 
\textit{Chaotic-Integrable Transition in the Sachdev-Ye-Kitaev Model},
Phys. Rev. Lett. {\bf 120}, 241603 (2018).

\bibitem{Shepelyansky17}
A.~R.~Kolovsky and D.~L.~Shepelyansky, 
\textit{Dynamical thermalization in isolated quantum dots and black holes},
Eur. Phys. Lett. {\bf 117}, 10003 (2017).

\bibitem{fn2}
Although the eigenvalues $\{\pm v_{i}\}$ of $J_{ij}$ are correlated, their sums, i.e.\ the eigenvalues of $\hat{H}_{2}$, become uncorrelated for large $N$.

\bibitem{fn1}
For two states $|n\rangle,|m\rangle$ we define the Hamming distance $|n-m|$ as
the number of bits in which the states differ. Containing four fermion creation/annihilation operators, and conserving fermion number parity, the matrix elements of the interaction
operator couple states of Hamming distance zero, two, and four.

\bibitem{fn3}
Here we ignore corrections of
$\mathcal{O}(\frac{1}{N})$.
However, for numerically accessible sizes
it is important to keep in mind the full expression
for the $H_4$ band width,
$\Delta_4 = \sqrt{\frac{3J^{2}}{4N^3}\binom{2N}{4}}$.

\bibitem{fn3a}
In order to compare the analytical predictions with numerical results without any fitting parameters it is important to use the full expression for the $H_{2}$ band width,
$\Delta_2 = \sqrt{\frac{\delta^{2}}{2N}\binom{2N}{2}}$.

\bibitem{RPmodel}
N.~Rosenzweig, C.~E.~Porter, \textit{Repulsion of energy levels in complex atomic spectra},
Phys. Rev. {\bf 120}, 1698 (1960).

\bibitem{Zirnbauer86} 
M.~R.~Zirnbauer, \textit{Localization transition on the Bethe lattice}, Phys. Rev. B {\bf 34}, 6394 (1986).


\bibitem{Lemarie17} 
I.~Garc\'ia-Mata, O.~Giraud, B.~Georgeot, J.~Martin, R.~Dubertrand, G.~Lemari\'e
\textit{Scaling theory of the Anderson transition in random graphs: ergodicity and universality}, 
Phys. Rev. Lett. {\bf 118}, 166801 (2017).

\bibitem{Lemarie20} 
I.~Garc\'ia-Mata, J.~Martin, R.~Dubertrand, O.~Giraud, B.~Georgeot, G.~Lemari\'e,
\textit{Two critical localization lengths in the Anderson transition on random graphs},
Phys. Rev. Research {\bf 2}, 012020 (2020).


\bibitem{NEE_SYK}
T.~Micklitz, F.~Monteiro, and A.~Altland,
\textit{Nonergodic Extended States in the Sachdev-Ye-Kitaev Model},
Phys. Rev. Lett. {\bf 123}, 125701 (2019).


\bibitem{Huse10}
A.~Pal and D.~A.~Huse,
\textit{Many-body localization phase transition},
Phys. Rev. B \textbf{82}, 174411 (2010).

\bibitem{Huse07}
V.~Oganesyan and D.~A.~Huse,
\textit{Localization of interacting fermions at high temperature},
Phys. Rev. B \textbf{75}, 155111 (2007).

\bibitem{Kravtsov2}
V.~E.~Kravtsov, B.~L.~Altshuler, and L.~B.~Ioffe,
\textit{Non-ergodic delocalized phase in Anderson model on Bethe lattice and
regular graph},
Annals of Physics {\bf 389}, 148 (2018). 

\bibitem{Biroli17}
G.~Biroli, and M.~Tarzia,
\textit{Delocalized glassy dynamics and many-body localization},
Phys. Rev. B {\bf 96}, 201114(R) (2017). 

\bibitem{Biroli12}
G.~Biroli, A.~C.~Ribeiro-Teixeira, and M.~Tarzia,
\textit{Difference between level statistics, ergodicity and localization
transitions on the Bethe lattice}, 
arXiv:1211.7334.

\bibitem{Kravtsov1}
B.~L.~Altshuler, E.~Cuevas, L.~B.~Ioffe, and V.~E.~Kravtsov,
\textit{Nonergodic Phases in Strongly Disordered Random Regular Graphs},
Phys. Rev. Lett. {\bf 117}, 156601 (2016).

\bibitem{DeLuca14}
A.~De Luca, B.~L.~Altshuler, V.~E.~Kravtsov and A.~Scardicchio,
\textit{Anderson Localization on the Bethe Lattice: Nonergodicity of Extended
States},
Phys. Rev. Lett. {\bf 113}, 046806 (2014).

\bibitem{fn8}
Notice that the inverse participation ratio here has not been normalized by
its value at $\delta=0$, as in our previous publication~\cite{NEE_SYK}.

\bibitem{Efetov}
K.~B.~Efetov, {\it Supersymmetry in Disorder and Chaos} (Cambridge Univ. Press, 1999).

\bibitem{crossing}
More specifically, we used $\delta_c = \frac{\sqrt{\pi} Z}{2 \sqrt{\rho}} \log(
\frac{\sqrt{\pi} Z}{32 \pi^2})$.

\bibitem{Ossipov}
K.~Truong and A.~Ossipov,
\textit{Eigenvectors under a generic perturbation: Non-perturbative results
from the random matrix approach},
Eur. Phys. Lett {\bf 116}, 37002 (2016).

\bibitem{fn6}
Unlike with low dimensional single particle problems, the
effectively high dimension of Fock space implies non-universality of the Thouless
energy. For example, non-zero mode corrections to the spectral form factor (the
Fourier transform of the two-point correlation function in energy) and the two-point 
function itself, respectively, become visible at different energy scales.

\bibitem{Bogomolny13}
Y.~Y.~Atas, E.~Bogomolny, O.~Giraud, and G.~Roux,
\textit{Distribution of the Ratio of Consecutive Level Spacings in Random
Matrix Ensembles},
Phys. Rev. Lett. {\bf 110}, 084101 (2013).

\bibitem{Zirn86}
M.R.Zirnbauer,
\textit{Anderson Localization and Nonlinear $\sigma$ Model With Graded
Symmetry},
Nucl. Phys. B [FS] \textbf{265}, 375 (1986).

\bibitem{hamming}
We here neglect the parametrically smaller number of sites with $|n-m|=2$ connected to $n$ by matrix elements changing the occupation of just two fermion orbitals.





\end{thebibliography}
\end{document}